\documentclass[a4paper,10pt]{article}
\usepackage[final]{graphicx}
\usepackage{url}
\usepackage{comment}
\usepackage{amsmath,amsfonts,bm,amssymb,tabularx,fancybox,ulem,mathtools,epsfig,epic,eepic,subfig,rotating}
\usepackage[usenames,dvipsnames,svgnames,table]{xcolor}
\usepackage{makeidx,dsfont,wrapfig}
\usepackage[colorinlistoftodos]{todonotes}
\usepackage{cool} 
\usepackage{paralist} 

\usepackage[english]{babel}

\usepackage{multirow}

\usepackage[linesnumbered,ruled,vlined]{algorithm2e}

\usepackage{tikz}
\usetikzlibrary{shapes,arrows}
\usetikzlibrary{
	decorations.pathreplacing,
	decorations.pathmorphing
}

\makeatletter
\def\BState{\State\hskip-\ALG@thistlm}
\makeatother

\usepackage{tikz}
\usetikzlibrary{shapes,arrows}
\usetikzlibrary{%
	decorations.pathreplacing,%
	decorations.pathmorphing%
}

\graphicspath{{./}{./figures/}{./figures/surf/}{./figures/conc/}{./conc/}{./surf/}{./water/}{./pol/}}

\excludecomment{omitext}

\newcommand{\n}{\noindent}

\newcommand{\bv}{{\bf v}}

\newcommand{\bx}{{\bf x}}

\newcommand{\bn}{{\bf n}}
\newcommand{\bK}{{\bf K}}

\newcommand{\dl}{\ensuremath{\partial}}
\newcommand{\p}[2]{\ensuremath{\frac{\partial #1}{\partial #2}}}
\newcommand{\lb}{\ensuremath{\lambda}}

\DeclareBoldMathCommand\balpha{{\alpha}}
\DeclareBoldMathCommand\btau{{\tau}}
\DeclareBoldMathCommand\bgamma{{\gamma}}
\DeclareBoldMathCommand\bbeta{{\beta}}
\DeclareBoldMathCommand\bgrad{{\nabla}}
\DeclareBoldMathCommand\bX{{\mathcal{X}}}

\def\pd1x#1{{\frac{\partial #1}{\partial x_1}}}

\newcommand{\qed}{\nobreak \ifvmode \relax \else
      \ifdim\lastskip<1.5em \hskip-\lastskip
      \hskip1.5em plus0em minus0.5em \fi \nobreak
      \vrule height0.75em width0.5em depth0.25em\fi}

\def\<-{\leftarrow }
\def\->{\rightarrow} 
\def\<={\Leftarrow }
\def\=>{\Rightarrow} 

\title{Modelling shear thinning polymer flooding using a dynamic viscosity model}
\author{Prabir ${\rm {Daripa}}^{\rm a}$\thanks{Author for correspondence (email:daripa@math.tamu.edu)}\ \ and\ \ Rohit ${\rm {Mishra}}^{\rm b}$\thanks{email:rmishra@tamu.edu}\\ 
${}^{\rm a}$Department of Mathematics\\
${}^{\rm b}$Mechanical Engineering\\
Texas A\&M University\\
College Station, TX 77843-3368}

\begin{document}

\maketitle

\begin{abstract}
Two distinct effects that polymers exhibit are shear thinning and viscoelasticity. The shear thinning effect is important as the polymers used in chemical enhanced oil recovery usually have this property. We propose a novel approach to incorporate this shear thinning effect through an  effective dynamic viscosity of the shear thinning polysolution. The procedure of viscosity calculation of the polysolution, although based on a very basic power law model, is based on empiric coefficients which depends on a spatio-temporally evolving variable namely concentration of polymer. Since viscosity calculation is done pointwise, the accuracy of the model is higher than what exists in literature. This method has been integrated with an existing method for a Newtonian physics based model of porous media flows. The solver uses a hybrid numerical method developed by Daripa \& Dutta~\cite{DFEMcode,daripa2017modeling,daripa2019convergence}. The above method
solves a system of coupled elliptic and transport equations modelling Darcy's law based polymer flooding process using a discontinuous finite element method and a modified method of characteristics. Simulations show (i) competing effects of shear thinning and mobility ratio; (ii) injection conditions such as injection rate and injected polymer concentration influence the choice of polymers to optimise cumulative oil recovery; (iii) permeability affects the choice of polymer; (iii) dynamically evolving travelling viscosity waves; and (v)  shallow mixing regions of small scale viscous fingers in homogeneous porous media. This work shows an effective yet easy approach to make design choices of polymers in any given flooding condition.

\end{abstract}

\n{\bf Article Highlights}
\begin{itemize}
    \item A dynamic viscosity framework for modeling enhanced oil recovery by  shear thinning polymer flooding is proposed.
    \item This framework is applied to an in-house surfactant-polymer flooding code.
    \item A new software that solves shear-thinning polymer flooding problem with or without surfactant has been developed.
\end{itemize}

\n {\bf Keywords}\ Shear thinning polymer flooding, Enhanced oil recovery, Power law fluids, Data driven modelling, Immiscible two-phase flow

\maketitle
\newpage
\section{\label{sec:introduction}Introduction}
Multiphase multicomponent porous media flows are ubiquitous. They occur in environment, climate, subsurface, biomedical sciences, etc. just to name a few. They also occur in chemical enhanced oil recovery (CEOR) by polymer and/or polymer-surfactant flooding, a topic related to the subject of this paper. In these chemical floods, a polymer-thickened aqueous phase which we will henceforth call polysolution is injected to displace oil. Assuming polysolution as a Newtonian fluid and neglecting capillary pressure, these flooding processes are modeled by a nonlinear system of elliptic equation for pressure and hyperbolic equations for pressure and saturation (see Daripa et al.~\cite{daripa1988polymer}). An increase in viscosity, due to polymer, of the displacing fluid inhibits fingering instability and increases the saturation behind the front sweeping the oil. In that study, it has been established through nonlinear wave analysis and numerical simulation using a front tracking method~\cite{dgglm86:reservoir} 
that both of these effects of polymer improve oil recovery. This method is not applicable as it is when capillary pressure is taken into account (in particular when surfactant is also used). 

In Daripa \& Dutta~\cite{daripa2017modeling}, authors extended the polymer model of Daripa et. al~\cite{daripa1988polymer} to include the effects of capillary pressure and also of another component, namely surfactant. This new model which we call SP-model (`S' stands for surfactant and `P' stands for polymer), is a coupled nonlinear system of elliptic and transport equations. This SP model and the numerical method are very briefly reviewed in \S \ref{subsec:Newtonian-model} and \S \ref{sec:numerical-method} which is adequate for the development of the remaining part of this paper. Convergence of this numerical method has been proven in Daripa \& Dutta~\cite{daripa2019convergence}. A Matlab code~\cite{DFEMcode} available on Github has been developed implementing this method.
Using this code, numerical simulations have been performed to validate this method qualitatively and quantitatively (see Daripa \& Dutta~\cite{daripa2019convergence}), and to evaluate the relative performance of several floods in homogeneous and heterogeneous reservoirs. As we will see below, this model and the numerical method are the building blocks for the data driven model and the method that incorporates the shear thinning effect of polymer. 
In this paper, we are interested in including non-Newtonian effects of polymer in SP-flooding model, develop a numerical method to numerically solve this new model, and then perform numerical simulations to validate this method. In particular, we perform simulations with polymer-flooding exclusively by setting the concentration of surfactant to zero in the SP model. This provides an understanding of data driven non-Newtonian effect of polymers on viscous fingering and oil recovery properties of polymer floods. In turn, as we will see, this is helpful in the selection of a polymer for maximum oil recovery at any given flooding condition. 

Polymers that are commonly used in chemical enhanced oil recovery are usually stiff polymers of shear thinning type. Thus, stability of such a displacement process where the displacing fluid is shear thinning plays an important role.
It is known from theoretical studies \cite{bonn1997viscoelastic,lindner2000viscous} that shear thinning behavior of non-Newtonian displacing fluids suppresses fingering instability. In these and many other studies, shear thinning fluids are generally modeled as power-law fluids which depend on two indices, shear rate and density of polysolution. These are taken as constants in linear stability studies. However, in practice such stability results are of little use in CEOR since power law indices are functions of the type of polymer and its concentration which change in space and time in the flow domain (see \S \ref{sec:mathematical-model} and \S \ref{sec:numerical-method} below). Stability of such a displacement process with time and space dependent parameters in the power law has not been studied to-date. Even if it were, such stability results are likely to be of little value for practical applications to CEOR since values of these parameters evolve in space-time and are not known a priori. For this and many applications related reasons, there is a need to better understand and model the displacement processes with non-Newtonian fluids in general. It appears that simulation is the only viable alternative for accurate prediction of flow features, flow instabilities and other oil recovery performance measures.

 
Towards this end, it is worth mentioning some other works from a huge body of literature on CEOR. CEOR methods increase oil recovery by altering fluid-fluid and/or fluid-rock interaction in the reservoir by reducing interfacial tension (IFT) and/or altering the viscosity of the injected fluid for mobility control. Another way the injected chemicals increase recovery is by altering the wettability of the rock to increase oil permeability \cite{zhou2018synthesis}\cite{zhou2015synthesis}. Pope et al. \cite{pope2011recent} reported that polymer flooding can increase oil recovery up to 12-30 \% original oil initially in-place (OOIP). The idea behind polymer flooding is to introduce an intermediate layer of higher viscosity fluid which increases the cumulative oil recovered by reducing mobility ratio and thereby reducing the finger formation of less viscous fluid (water) displacing the high viscous fluid (oil). Although this method has been used widely in industry, it still lacks a fundamental understanding of the role of polymers in CEOR. Sheng et al.~\cite{sheng2015status} in their review paper on the status of polymer flooding listed different challenges associated with polymer flooding. The major challenge is associated with the basic underlying physics of polymer and its behavior at variable rheological conditions such as temperature, salinity, permeability among others. Due to this highly variable nature of polymers it becomes a challenge in predicting oil recovery by simulation or by experiments. 

A polymer working in a specific set of conditions might not be viable in a different oil field completely. The complexity around usage of polymer is so profound that there is no general consensus on a very basic injection parameter, namely, the amount of polymer injected. Between 1970s to 1980s the amount of polymer injected is about 100-200 ppm PV (pore volume) in Chinese projects which changed to 500 to 600 ppm PV in early 1990s. This changed again in 2000s to 400-500ppm PV \cite{sheng2010modern}.  Niu et al. \cite{niu2006research} reports that when the amount of polymer injected is larger than 400ppm PV, incremental oil recovery becomes less sensitive to the amount of polymer injected. Levitt et al. \cite{levitt2013polymer} finds a similar trend,  namely increasing polymer (HPAM) viscosity from 3 cP to 60 cP does not significantly change recovery from two pore volumes (PV) of tertiary polymer injection. Apart from the above mentioned uncertainty, another reason polymer flooding may not be the ideal flooding technique is the higher injection rate required to inject polymers. This is overcome by yet another advanced CEOR method termed as the Surfactant-Polymer (SP) flooding. SP flooding is a chemical oil recovery method that introduces a layer of surfactant before injecting the polymer which mobilizes the oil by reducing the capillary pressure. However, to better understand the non-Newtonian effects of polymers we are only focusing on shear thinning polymer flooding in this paper.

Prior  research  conducted  in  the  field  of  Saffman-Taylor  instabilities  suggests that instabilities arise when a less viscous fluid displaces a fluid with higher viscosity.  It has been seen from theoretical studies \cite{lindner2000viscous,bonn1997viscoelastic} that shear thinning behavior of non-Newtonian displacing fluids may minimize the instability effects (finger formation) on the interface separating the displaced  and  displacing  fluids. The polymers that are currently used in field operations are considered inelastic shear thinning fluids. There have been numerical studies analyzing the inelastic behavior of polymers. The study by Durst et al.~\cite{durst1982laminar} shows that elastic behavior is observed only above a critical Deborah (De) number. This number is very small when we  consider velocity of shear thinning fluid in porous medium. This has been shown experimentally by Marshall \& Metzner~\cite{marshall1967flow}. Therefore, the focus of efforts in the past has been to analyze the shear thinning behavior of non-Newtonian fluids as the elastic behavior has less significance in this field of application.

After establishing the importance of shear thinning effects of polymer, the next big question is which polymer is the best for oil recovery. While there is no one good answer to this question multiple studies indicate a couple of good choices. The choice of polymer plays a crucial role in CEOR. The benefits of polymer flood is contingent on the extent of polymer retention in the oil field with minimum retention favorable for high recovery~\cite{sorbie1991introduction}. The three major mechanisms identified by Willhite et al.~\cite{willhite1977mechanisms} for polymer flow in porous media are polymer adsorption, mechanical entrapment and hydrodynamic retention.

Apart from the performance of polymer in oil recovery, the choice must also be made based on its environmental impacts. Natural polymers such as Xanthane and Schizophyllan are less detrimental to the environment when compared to hydrolyzed polyacrylamide (HPAM), a widely used industrial polymer which can cause environmental problems (increases the difficulty in oil-water separation, degrades naturally to produce toxic acrylamide and endanger local ecosystem). Agi et al.~\cite{agi2018natural} reviewed different natural polymers in reference to its application in enhanced oil recovery. Many previous studies show non-Newtonian behavior of natural polymer. For example,  Liu et al.~\cite{liu2014study} in their rheology study of Cassava starch finds that the polymer exhibits shear thinning behavior i.e. viscosity decreases with increasing strain rate. 
Most of the other natural polymers show similar behavior. These natural polymers in an eco-friendly way enhances oil recovery with higher sweep efficiency. This sweep efficiency can be improved if these are used as nanofluids \cite{agi2018mechanism}. Nanoparticles have advantages of being tolerant to high salinity, high temperature and retention in highly permeable reservoir. However, there are concerns on the cost of full-scale field implementation and toxicity of nanofluids. Agi et al~\cite{agi2020application} identified Cissus populnea(CP) as an important biopolymer and showed an efficient process to synthesize Cissus populnea nanofluid (CPNF). They showed at same concentrations the viscosity of CPNF is higher than CP which in turn is higher than commonly used natural polymer – Xanthane. While Xanthane shows a decrease in viscosity with increasing temperatures, CP and CPNF show an evident increase in viscosity. All these recent studies show promise on the possibility of improving enhanced oil recovery if the properties of the polymers are understood and applied carefully. Introduction of biopolymer nanoparticles is another emerging field where focus must be given to analyze the fluid properties not just experimentally but through polymer flood simulations.

Extensive numerical and computational studies for tertiary oil recovery is available in the literature. Daripa et al.~\cite{daripa1988polymer} provids a comprehensive analytical study of instability control in tertiary oil recovery. Afsharpoor et al.~\cite{afsharpoor2014micro} uses upper convected Maxwell equation to model strong extensional flow effect to relate the shear stress with the shear rate. Clemens et al.~\cite{clemens2013pore} performs a pore-scale evaluation of polymers displacing oil using experiments and CFD simulation. The experiments show shear thinning behavior of the polymer. CFD simulations show that viscosity is lower at pore throats than that in the pore. This shear thinning behavior affects the displacement efficiency of the polymer flooding. However, the power law model with fixed parameters $(n, \epsilon)$ is usually considered in CFD simulations when in relality these should be functions of polymer concentration. This assumption of constant parameters becomes even more detrimental in predicting recovery in real flooding simulations as the concentration of the polymer changes in space and time. Nandwani et al.~\cite{nandwani2019chemical} models surfactant flooding process using ANSYS FLUENT (a commercial CFD code) and shows that ultralow IFT, minimal fingering and low diffusion rate of the surfactant in oil are responsible for higher oil recovery. For Newtonian fluids, Daripa~\cite{daripa2008studies} studies a system of three fluids where the driving fluid drives the fluid in the middle which simultaneously drives the third fluid. He shows that there exists a critical viscosity for the middle fluid which leads to the most stable setup i.e. least finger growth. Recently, Manzoor et. al.~\cite{manzoor2020modeling} tested the effects of pressure variations on heavy oil recovery process in a homogeneous glass beads-packed physical model.  The study compares the simulation results with the experiments. The main conclusions are that maximum pressure and periodic pressure variations have the potential to enhance heavy-oil recovery. Moreover, they report that the numerical simulations are highly sensitive to uncertainty in permeability, porosity, diffusion coefficient of polymer, number of grid points and heavy-oil density.

\begin{figure}[ht!]
 \centering
 
  \includegraphics[width=8cm, height=8cm]{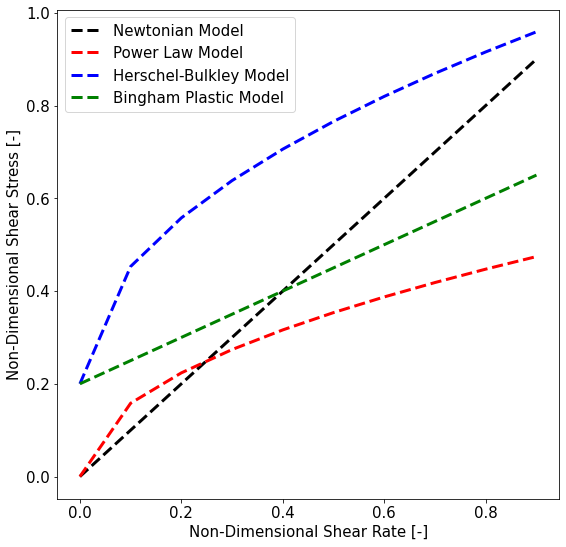}
  \caption{Rheological models}
  
  \label{fig:fig2}
\end{figure}

Xie et al. \cite{xie2021self} perform experiments on a microchip with heterogeneous porous structures where oil is displaced by dispersed polymer. They show dispersion effect even when the polymer particles are smaller than the pore throat. So the plugging effect is not the major mechanism for preferential flow control by dispersed polymers. Simulations \cite{xie2021self}\cite{xie2018lattice} for this type of flow show that dispersed polymers smartly controls the preferential flow by inducing pressure fluctuations and thereby increasing efficiency. Xie et al. \cite{xie2016lattice} uses Herschel-Bulkley model to find viscosity of the polymer in a two phase (polymer aqueous solution and oil) Lattice-Boltzman simulation. They show that apart from modelling shear rheological behavior of polymer it is also important to model the three phases (polymer, water and oil) individually. Although the Herschel-Bulkley model has been widely used in the literature to model pseudo-plastics, we use the power law model. The major difference between the two is that the power law model has one less empirical parameter. The reason for this choice is the widespread availability of the power-law coefficients for different polymers used in this study. The different rheological models important for polymer flooding simulations are shown in figure \ref{fig:fig2}.

Surfactant-polymer flooding simulations helps us in understanding the complex phenomenon of multi-component fluids moving across a porous region.  The simulations get even more challenging as one of the components (polymer) behaves as a non-Newtonian fluid i.e.  shear rate is not a linear function of strain rate.  To account for this, we propose and numerically study a data driven approach to incorporate shear thinning effect of polymer (shear thickening case is similar) in a hybrid numerical method developed by Daripa \& Dutta~\cite{daripa2017modeling,daripa2019convergence}. 
The data driven approach proposed here is based on the recognition that (i) shear thinning property can be modeled by an effective viscosity of polysolution based on a strain rate using a power law model, (ii) values of the power law parameters in the power law model that fit for the given shear rate and concentration of polymer vary and can be  estimated from curve fit based on experimental results, (iii) effective viscosity can be used in an otherwise Darcy's law based Newtonian model of SP floods without changing the fundamental equations yet implementing an accurate physical model which reduces to the Newtonian rheology with a constant viscosity in regions where the polymer concentration is zero, and (iv) shear thinning effect of any displacing fluid is easily incorporated in any existing Darcy's law and  Newtonian fluid mechanics based code. This paper is laid out as follows. Data driven mathematical model of SP flooding, which consists of the governing equations for SP flooding based on Newtonian fluid dynamics and Ostwald de Waele power law model for shear thinning polysolution, is described in \S \ref{sec:mathematical-model}. Numerical method is described in \S \ref{sec:numerical-method}. Numerical results in rectilinear and quarter five-spot geometries are presented in \S \ref{sec:results}. Finally we conclude in \S \ref{sec:conslusions}. The algorithm and the flow chart for the method are given in Appendix~\ref{appendix:algorithm}. The method for calculation of finger width is given in Apprndix~\ref{appendix:fingerwidth-calculation}.

\section{\label{sec:mathematical-model}Non-Newtonian Mathematical model}
The non-Newtonian mathematical model is built from combining Newtonian model of porous media flow for multi-phase multi-component porous media flow involving polymer as a component among others with power law model for the shear thinning fluid. This is discussed in this section. Next section discuses the numerical method.

\subsection{Newtonian model  of SP flooding}\label{subsec:Newtonian-model}
The Newtonian model of Surfactant-Polymer (SP) flooding in porous media is used here as a prototype model. In this section, we recall from Daripa \& Dutta~\cite{daripa2017modeling} some relevant facts about this model. It is given by the following coupled nonlinear system of elliptic equation $\eqref{eq:presssure}_2$ for global pressure $p$ and transport equations~\eqref{eq:saturation}, \eqref{eq:polymer} and \eqref{eq:surfactant} for saturation $s$, concentration $c$ of polymer, and concentration $\Gamma$ of surfactant respectively. 
\begin{equation}\label{eq:presssure}
\bv = -\bK(\bx)\lambda(s,c,\Gamma){\bf\nabla} p,\quad -{\bf\nabla}\cdot\left(\bK(\bx)\lambda(s,c,\Gamma){\bf\nabla} p\right) = q_a + q_o,
\end{equation}
\begin{equation}\label{eq:saturation}
\phi \frac{\partial s}{\partial t} + \frac{\partial f_a}{\partial s}{\bv}\cdot{\bf\nabla} s + {\bf\nabla}\cdot\left(D \frac{\partial p_c}{\partial s}{\bf\nabla} s\right)  \\ = g_s - \frac{\partial f_a}{\partial c}{\bv}\cdot{\bf\nabla} c - \frac{\partial f_a}{\partial \Gamma}{\bv}\cdot{\bf\nabla}\Gamma  - {\bf\nabla}\cdot\left(D\frac{\partial p_c}{\partial \Gamma}{\bf\nabla}\Gamma\right),
\end{equation}
\begin{equation}\label{eq:polymer}
\phi \frac{\partial c}{\partial t} + \left(\frac{f_a}{s}{\bv}+ \frac{D}{s} \frac{\partial p_c}{\partial s}{\bf\nabla} s + \frac{D}{s} \frac{\partial p_c}{\partial \Gamma}{\bf\nabla} \Gamma\right)\cdot{\bf\nabla} c    = g_c, 
\end{equation}
\begin{equation}\label{eq:surfactant}
\phi \frac{\partial \Gamma}{\partial t} + \left(\frac{f_a}{s}{\bv}+ \frac{D}{s} \frac{\partial p_c}{\partial s}{\bf\nabla} s + \frac{D}{s} \frac{\partial p_c}{\partial \Gamma}{\bf\nabla} \Gamma\right)\cdot{\bf\nabla}\Gamma   = g_\Gamma.
\end{equation}
The terms in the above equations are defined in Table-1 and after Table-1.

\begin{table}[h!]
\centering
 \begin{tabular}{||c c||} 
 \hline
 Term & Description  \\ [0.5ex] 
 \hline\hline
 $\bK(\bx)$ & Absolute Permeability Tensor\\ 
 \hline
$\lambda(s,c)$ & Total Mobility\\
 \hline
 $q_a , q_o$ & Source/sink terms for global pressure equation\\
 \hline
 $s, c, \Gamma$ & Saturation (volume fraction), Concentration of polymer, Concentration of surfactant\\
 \hline
 $\phi$ & Porosity of the medium\\
 \hline
 $p_c$ & Capillary pressure\\
 \hline
  $k_{rj}$ & (j=o) Relative permeability of oil; (j=a) Relative permeability of aqueous phase\\
 \hline
  $\mu_{j}$ & (j=o) Viscosity of oil (constant); (j=a) Viscosity of aqueous phase\\
  \hline
  $f_{j}$ & (j=o) fractional flow function of oil; (j=a) fractional flow function of the aqueous phase\\
 \hline
  $g_s,g_c,g_\Gamma$ & Source terms for saturation, concentration and surfactant transport
 \\ [1ex] 
 \hline
\end{tabular}\\
\caption{Nomenclature}
\label{table:1}
\end{table}

The elliptic equation~$\eqref{eq:presssure}_2$ solves for a new function called global pressure which is defined in terms of a thermodynamic pressure and capillary phase pressures so that we have a canonical elliptic equation \eqref{eq:presssure} for the global pressure $p$. The following relation between this global pressure and phase pressures is taken from Daripa \& Dutta~\cite{daripa2017modeling}. 
\begin{align}\label{gblpnew}
p = \frac{1}{2} (p_o+p_a) &+ \frac{1}{2} \int^{s}_{s_c}(f_o-f_a)(\zeta,c,\Gamma)\frac{dp_c}{d\zeta}(\zeta,\Gamma)d\zeta 
+ \frac{1}{2} \int^{\Gamma}_{\Gamma_c}(f_o-f_a)(s,c,\xi)\frac{dp_c}{d\xi}(s,\xi)d\xi \nonumber \\
&- \frac{1}{2}\int \eta^c(s,c,\Gamma) \left(\p{c}{x}dx +\p{c}{y}dy\right)- \frac{1}{2}\int \eta^s(s,c,\Gamma) \left(\p{s}{x}dx +\p{s}{y}dy\right)\nonumber \\
&- \frac{1}{2}\int \eta^{\Gamma}(s,c,\Gamma) \left(\p{\Gamma}{x}dx +\p{\Gamma}{y}dy\right) + C,
\end{align}
where $C$ is a reference pressure which takes the place of an integration constant in the calculations. In the above, 
\begin{subequations}
	\begin{align}
	\eta^c(s,c,\Gamma) &=\int^{s}_{s_c}\frac{\dl}{\dl c}\left(f_o-f_a\right)(\zeta,c,\Gamma)\frac{dp_c}{d\zeta}(\zeta,\Gamma)d\zeta + \int^{\Gamma}_{\Gamma_c}\frac{\dl}{\dl c}\left(f_o-f_a\right)(s,c,\xi)\frac{dp_c}{d\xi}(s,\xi)d\xi, &\label{gblpnew-1}\\
	\eta^s(s,c,\Gamma) &=\int^{\Gamma}_{\Gamma_c}\frac{\dl}{\dl s} \left(f_o-f_a\right)(s,c,\xi)\frac{dp_c}{d\xi}(s,\xi)d\xi, &\label{gblpnew-2}\\
	\eta^{\Gamma}(s,c,\Gamma) &=\int^{s}_{s_c}\frac{\dl}{\dl \Gamma}\left(f_o-f_a\right)(\zeta,c,\Gamma)\frac{dp_c}{d\zeta}(\zeta,\Gamma)d\zeta, & \label{gblpnew-3}
	\end{align}
\end{subequations}
where $s_c$ is the value of the aqueous phase saturation at which $p_c(s_c,\Gamma) = 0$ and $\Gamma_c$ is the surfactant concentration value defined similarly by $p_c(s,\Gamma_c) = 0$.

The aqueous phase viscosity $\mu_a$ in this Newtonian model depends linearly on $c$: $\mu_a=\mu_w+\beta c$ where $\mu_w$ is the constant viscosity of water and $\beta$ is a polymer specific constant. Thus, this system does not include non-Newtonian effect of polymer. The phase relative permeability $k_{rj}$ depends on saturation $s$ and capillary pressure dependent residual saturation $s_r$ (see \cite{daripa2017modeling}). Since capillary pressure depends on concentration $\Gamma$ of surfactant, the phase relative permeability $k_{rj}$ depends on both $s$ and $\Gamma$. The total mobility $\lambda = \lambda_a + \lambda_o$ where $\lambda_{j} =  k_{rj}/\mu_{j}$ is the mobility of phase $j$. Therefore $\lambda_{a}$ depends on the triplet ($s,c,\Gamma$) but $\lambda_{o}$ depends only on ($s,\Gamma$). However, both of these mobility we write generically as $\lb_j(s,c,\Gamma)$ for convenience. The function $f_j(s,c,\Gamma) = \lb_j(s,c,\Gamma)/\lb(s,c,\Gamma)$ is the fractional flow function of the phase $j$ and the coefficient $D(s,c,\Gamma) =  \bK(\mathbf{x})\lb_o(s,\Gamma)f_a(s,c,\Gamma)$. 

The source/sink terms in the elliptic equation~$\eqref{eq:presssure}_2$ are given by (see Daripa \& Dutta~\cite{daripa2017modeling})
\begin{align}\label{source-term}
&q_a =\begin{cases}
Q \\
-(\lb_a/\lb) \, Q \, ; \\
0 
\end{cases} 
q_o = \begin{cases}
0   \\
-(\lb_o/\lb) \, Q  \\
0 
\end{cases} 
\text{ at} \,
\begin{cases}
\bx^i = (0,0) \qquad \qquad \qquad \text{ (Source)} \\
\bx^p = (1,1) \qquad \qquad \qquad \text{ (Sink)} \\
\bx \in \Omega\setminus\{(0,0) \cup (1,1)\} \, \text{ (Elsewhere)}
\end{cases},
\end{align}
and
three source terms in the transport equations are defined as (see Daripa \& Dutta~\cite{daripa2017modeling})
\begin{align*}
g_s =\begin{cases}
(1-f_a)Q\\
0 
\end{cases} 
g_c = \begin{cases}
(c^i - c) Q/s \\
0 
\end{cases}
g_\Gamma = \begin{cases}
(\Gamma^i - \Gamma) Q/s \\
0 
\end{cases}
\text{ at } \bx = \begin{cases}
\bx^i \qquad \quad \text{ (source)}\\
\Omega \setminus \{\bx^i\} \text{ (elsewhere)}
\end{cases},
\end{align*}
where $Q$ is the volumetric injection/production rate. 
The following initial and boundary conditions are prescribed.
\begin{subequations}
	\begin{align}
	\quad s(\bx) = s_0(\bx), \quad  c(\bx) = c_0(\bx), \quad \Gamma(\bx) = \Gamma_0(\bx); \quad & \bx \in \Omega, \quad t = 0, \label{ic}\\
	\quad \bgrad s\cdot \hat{\bn} = 0, \quad  \bgrad c\cdot \hat{\bn}  = 0, \quad \bgrad \Gamma\cdot \hat{\bn}  =0 \quad \& \quad  \bv_j\cdot\hat{\bn} = 0; \qquad & \bx \in \dl \Omega, \quad t > 0, \quad (j = a,o), \label{bc}
	\end{align}
\end{subequations}
where $\hat{\bn}$ is a unit vector normal to $\dl \Omega$.

This SP-flooding model reduces to polymer flooding model if we set surfactant concentration $\Gamma=0$ in the above equations. This eliminates the transport equation~\eqref{eq:surfactant} for surfactant, the source term $g_\Gamma$, the terms involving $\Gamma$ from equations~\eqref{eq:saturation} and \eqref{eq:polymer}, and dependency of $D(s,c,\Gamma)$, $\lambda(s,c,\Gamma)$ and fractional flow functions $f_j(s,c,\Gamma)$ on $\Gamma$. This completes the description of the Newtonian model of Surfactant-Polymer (SP) flooding. Finer details can be found in Daripa \& Dutta~\cite{daripa2017modeling}.

\subsection{\label{sec:shear thinning}Model of shear thinning}
To include shear thinning effect of polymers in the above model, following Ostwald de Waele power law model~\eqref{eq:powerlaw-model} is used for the  calculation of the effective viscosity $\mu_a$ of the aqueous phase. 
\begin{equation}\label{eq:powerlaw-model}
\mu_a=\rho\varepsilon \dot{\gamma}^{n-1},
\end{equation}
where shear rate $\dot{\gamma}$ is given by 
\begin{equation}\label{eq:strainrate}
\dot{\gamma}=2\sqrt{\mid\Pi_D\mid},
\end{equation}
in terms of the second invariant $\Pi_D$ of the strain rate tensor given by~\cite{schobeiri2014applied}
\begin{equation}\label{eq:straintensor}
\Pi_D=-\frac{1}{4}\Bigg[\Bigg(\frac{\partial u}{\partial y}+\frac{\partial v}{\partial x}\Bigg)\Bigg]^2+\frac{\partial u}{\partial x}\frac{\partial v}{\partial y}.
\end{equation} 
The power law~\eqref{eq:powerlaw-model} depends on density $\rho$, strain rate $\dot\gamma$, and two parameters $(\varepsilon, n)$. The density $\rho(\bf{x},t)$ is the weighted sum of water and polymer densities based on the local in time and space value of concentration, $c(\bf{x},t)$, of polymer which evolves according to its transport equation~\eqref{eq:polymer}. The strain rate $\dot{\gamma}$ relates to the second invariant of the strain rate tensor, $\Pi_D$, according to \eqref{eq:strainrate}~\cite{schobeiri2014applied}. 

\begin{figure}[ht!]
\begin{minipage}{\textwidth}
\centering
\renewcommand{\tabcolsep}{0.08cm}
\begin{tabular}{ccccc}

\includegraphics[width=2.5in]{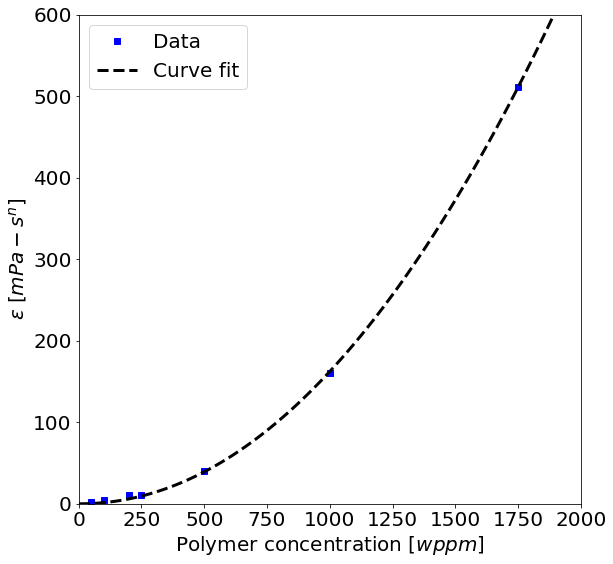}&
\includegraphics[width=2.5in]{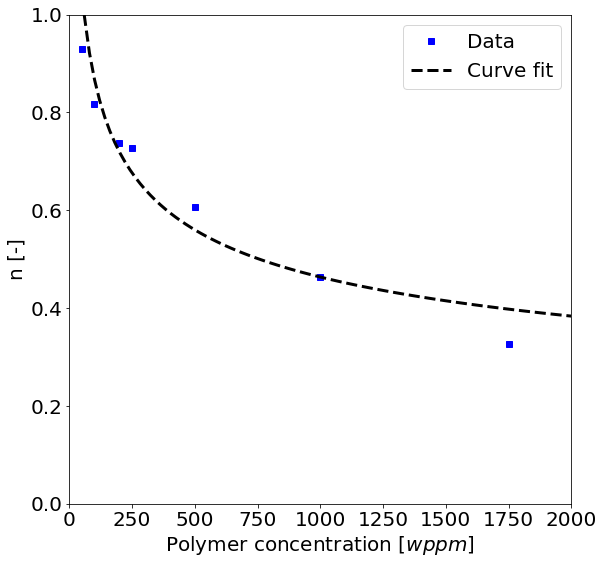}\\
\footnotesize{$\mathbf{(a)}$}&\footnotesize{$\mathbf{(b)}$}\\
\includegraphics[width=2.5in]{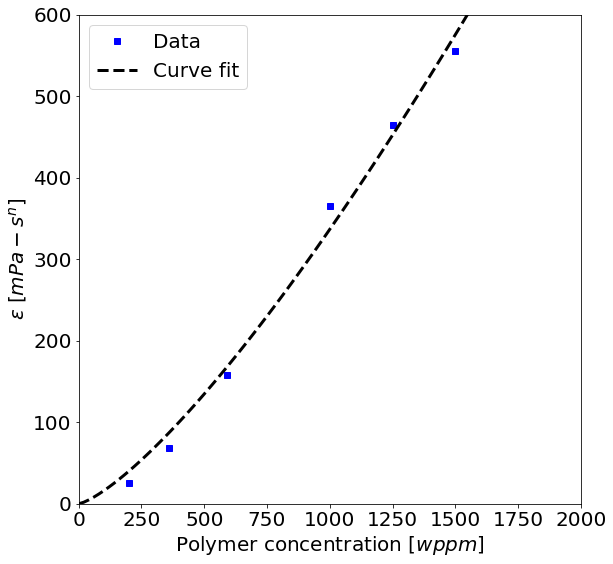}&
\includegraphics[width=2.5in]{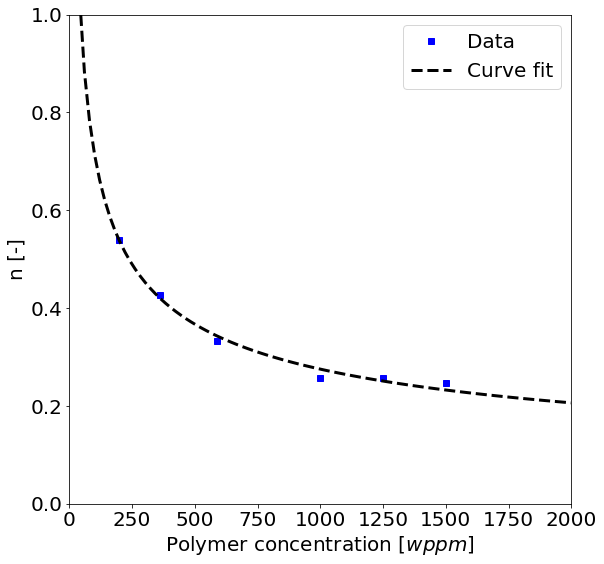}\\
\footnotesize{$\mathbf{(c)}$}&\footnotesize{$\mathbf{(d)}$}\\
\end{tabular}
\medskip
\caption{Top row: $\varepsilon$ vs concentration for Xanthane (L) $n$ Vs concentration for Xanthane (R)\cite{lindner2000viscous}\\
Bottom row: $\varepsilon$ vs concentration for Schizophyllan (L) $n$ vs concentration for Schizophyllan (R)\cite{hatscher2016schizophyllan}} 
\label{fig:fig3}
\end{minipage}
\end{figure}

The strain rate $\dot{\gamma}$ is calculated from the velocity field $(u,v)=\bv$ which is obtained as part of the solution of model~\eqref{eq:presssure} through \eqref{eq:surfactant} described in the next section~\S \ref{sec:numerical-method}. The parameters $(\varepsilon, n)$ depend on the type and concentration of polymers. This is exemplified in Fig.~\ref{fig:fig3} in which experimental values of $\varepsilon$ and $n$ versus concentration for two shear thinning polymers, namely Xanthane and Schizophyllan, are shown from Hatscher~\cite{hatscher2016schizophyllan} and Lindner et al.~\cite{lindner2000viscous}.
Values of these two parameters in the flow domain at any specific time are found from curve fitting local values of $c$ with experimental data similar to the ones shown in Fig.~\ref{fig:fig3} for the specific polymers in use. Local values of density $\rho$ of polysolution is also found from local values of $c$, concentration of polymer. Similarly, values of shear rate $\dot{\gamma}$ in the flow domain at any specific time are found from velocity field $(u,v)=\bv$ which is obtained as part of the solution of the model \eqref{eq:presssure} through \eqref{eq:surfactant} described in \S \ref{sec:numerical-method}. All these values are then used in the power law~\eqref{eq:powerlaw-model} to find the local effective viscosity $\mu_a(\bx,t)$. This affects the model equations~\eqref{eq:presssure}-\eqref{eq:surfactant} through the  functions $\lambda_a, \lambda, f_a$ and $D$ appearing in these equations which depend on $\mu_a(\bx,t)$. This procedure is a significant improvement and provides accurate values of interest than conventional practices in which usually power law parameter values are taken at fixed values of polymer concentration at the injection point. In \S\ref{sec:numerical-method}, we show this by integrating shear thinning model in the numerical method.

This nonlinear coupling of effective viscosity $\mu_a(\bx,t)$ with the system~\eqref{eq:presssure}-\eqref{eq:surfactant} and influence of experimental data make it increasingly hard to predict associated response of viscosity variation in the flow domain, flow features, fingering instability and oil recovery performance measures to shear thinning. Therefore, accurate physical interpretation of numerical solution of this data driven model problem is not going to be easy. Regardless, various numerical results obtained using this data driven numerical method discussed in the next section are analyzed and interpreted later in terms of physics. 



\section{\label{sec:numerical-method}Numerical Method}

\subsection{\label{sec:Newtonian-numerical-method}A Brief Review of the Numerical Method for the Newtonian Model}

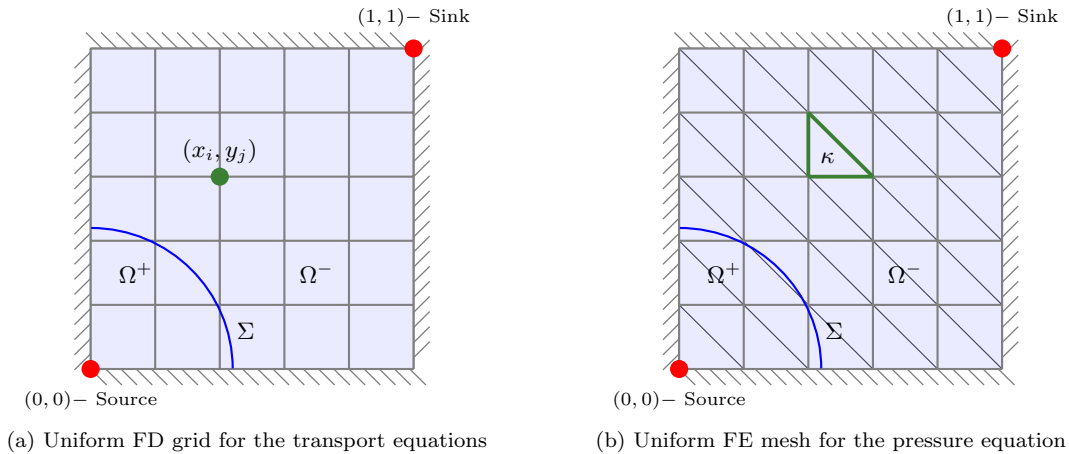
\begin{figure}[h!]
	\centering
	\subfloat[Uniform FD grid for the transport equations]{
	\begin{tikzpicture}[scale=0.85,
	interface/.style={
		postaction={draw,decorate,decoration={border,angle=-45,
				amplitude=0.3cm,segment length=2mm}}},]
	\draw (0,5) -- (5,5);
	\draw (5,0) -- (5,5);
	\draw (0,0)-- (0,5);
	\draw (0,0)--(5,0);
	
	\fill[blue!25!,opacity=.3] (0,0) rectangle (5,5);
	\draw [thick, gray] (0,0) grid (5,5);	
	
	\draw[blue,thick] (0:2.2) arc (0:90:2.2) ;	
	
	\draw	(0,-0.2) node [text=black,below] {\scriptsize $(0,0)-$  Source}
	(5,5.2) node [text=black,above] {\scriptsize $(1,1)-$ Sink}
	(3.5,1.5) node [text=black] {\small $\Omega^-$}
	(.7,1.5) node [text=black] {\small $\Omega^+$}
	(2.4,.6) node [text=black] {\small $\Sigma$}
	(2,3.4) node [text=black,thick] {\small $(x_i,y_j)$};
	
	\draw[gray,line width=.5pt,interface](0,0)--(5,0) (5,0)--(5,5) (5,5)--(0,5) (0,5)--(0,0);
	
	\foreach \Point in {(0,0),(5,5)}{\fill[red] \Point circle[radius=4pt];}
	\foreach \Point in {(2,3)}{\fill[OliveGreen] \Point circle[radius=4pt];}
	
	\end{tikzpicture} }\qquad \qquad
\subfloat[Uniform FE mesh for the pressure equation]{
	\begin{tikzpicture}[scale=0.85,
	interface/.style={
		postaction={draw,decorate,decoration={border,angle=-45,
				amplitude=0.3cm,segment length=2mm}}},]
	\draw (0,5) -- (5,5);
	\draw (5,0) -- (5,5);
	\draw (0,0)-- (0,5);
	\draw (0,0)--(5,0);
	\draw (0,1) -- (1,0);
	\draw (0,2) -- (2,0);
	\draw (0,3) -- (3,0);
	\draw (0,4) -- (4,0);
	\draw (0,5) -- (5,0);
	\draw (1,5) -- (5,1);
	\draw (2,5) -- (5,2);
	\draw (3,5) -- (5,3);
	\draw (4,5) -- (5,4);

	\fill[blue!25!,opacity=.3] (0,0) rectangle (5,5);
	\draw [thick, gray] (0,0) grid (5,5);

	\draw[blue,thick] (0:2.2) arc (0:90:2.2) ;	
	
	\draw	(0,-0.2) node [text=black,below] {\scriptsize $(0,0)-$  Source}
	(5,5.2) node [text=black,above] {\scriptsize $(1,1)-$ Sink}
	(3.5,1.5) node [text=black] {\small $\Omega^-$}
	(.7,1.5) node [text=black] {\small $\Omega^+$}
	(2.4,.6) node [text=black] {\small $\Sigma$}
	(2.3,3.3) node [text=black,thick] {\small $\kappa$};	
	
	\draw[gray,line width=.5pt,interface](0,0)--(5,0) (5,0)--(5,5) (5,5)--(0,5) (0,5)--(0,0);
	
	\foreach \Point in {(0,0),(5,5)}{\fill[red] \Point circle[radius=4pt];}
	
	\draw[line width=0.5mm, OliveGreen] (2,3) -- (2,4);
	\draw[line width=0.5mm, OliveGreen] (2,4) -- (3,3);
	\draw[line width=0.5mm, OliveGreen] (3,3) -- (2,3);
	
	\end{tikzpicture}}
	\caption{Discretization of the computational domain for (a) the Transport Equations and (b) the Pressure Equation.}

	\label{fig:grid}
\end{figure}

We first briefly review from Daripa \& Dutta~\cite{daripa2017modeling} the hybrid numerical method for solving the mathematical model of \S \ref{sec:mathematical-model}. In this method, global pressure equation \eqref{eq:presssure} is solved using a {\underline D}iscontinuous {\underline F}inite {\underline E}lement {\underline M}ethod (DFEM)~\cite{HWW2010}. The system of transport equations~\eqref{eq:saturation}, \eqref{eq:polymer}, \& \eqref{eq:surfactant} is solved by a time-implicit finite difference method based on the {\underline M}odified {\underline M}ethod {\underline O}f {\underline C}haracteristics (MMOC)~\cite{DR1982,D1983}. Two different types of grid are used in the numerical method: the finite difference grid shown in Fig.~\ref{fig:grid}(a) for the transport equations and the finite element grid shown in Fig.~\ref{fig:grid}(b) for the pressure equation. Thus, transfer of data between these two grid system during numerical solution process is required in this numerical method because of the coupling of elliptic and transport equations. Details about the computational grid, the numerical method, and the entire computational algorithm for the above Newtonian SP flooding model are given in Daripa \& Dutta~\cite{daripa2017modelling}. The convergence of the method has been proven in one-dimension in Daripa \& Dutta~\cite{daripa2019convergence}.

\subsection{\label{sec:non-Newtonian-numerical-method}Numerical Method for the non-Newtonian Model}
The numerical method discussed above in \S \ref{sec:Newtonian-numerical-method} is adapted for this shear thinning model by first calculating the aqueous phase (polysolution) viscosity $\mu_a(\bx,t)$ using data driven power law model~\eqref{eq:powerlaw-model} as discussed in \S \ref{sec:shear thinning}. The effective viscosity and appropriate associated quantities which depend on this viscosity are calculated at all grid points of both types of grid. For example, total mobility $\lambda$ which the elliptic equation~$\eqref{eq:presssure}$ depends on is calculated at all finite element grid points at every time level since effective viscosity $\mu_a$ changes in time which $\lambda$ depends on as explained in \S \ref{sec:shear thinning}. Similarly, $\lambda_a, \lambda, f_a$ and the function $D$ which the transport equations~\eqref{eq:saturation}, \eqref{eq:polymer}, \& \eqref{eq:surfactant} depend on are calculated at all finite difference grid points at every time level. All this adds to the computational cost to study the shear thinning effect of polymer. In appendix~\ref{appendix:algorithm}, the entire algorithm that implements this numerical method and the flow chart are given. 

\section{\label{sec:results}Results} 
Three sets with two simulations per set (for two polymers) are carried out for homogeneous and two different heterogeneous permeability fields. Extensive comparisons have been made to validate this approach. Results are summarised below. 

\subsection{Shear thinning induced nonuniform viscosity and travelling waves}
As discussed in \S \ref{sec:shear thinning}, variable viscosity profile in a flow domain due to shear thinning effect is very hard to predict a priori due to the influence of experimental data and permeability field of the porous media. It is even harder to predict its development in time a priori as the aqueous phase viscosity $\mu_a$, calculated in space and time from the shear thinning power law model~\eqref{eq:powerlaw-model}, is nonlinearly coupled with the elliptic equation for pressure and the transport equations for saturation and polymer as discussed earlier. However, a posteriori one can analyze such profiles to get an insight into the development of any phenomena and/or distinct patterns discovered in the process. Here we exemplify this by analyzing the evolution of such profiles in a quarter five spot simulation in heterogeneous porous media. 

\begin{figure}[ht!]
 \centering
 
  \includegraphics[scale=0.4]{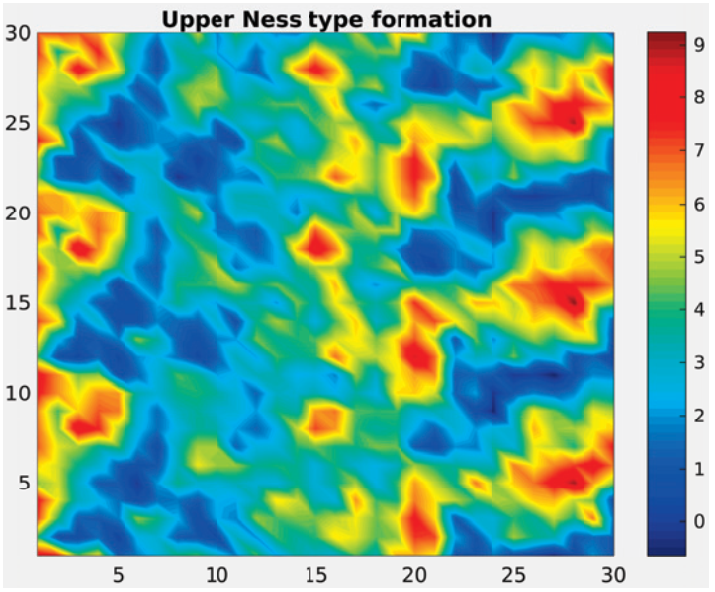}
  \caption{Logarithmic permeability plot from the SPE10 benchmark dataset on a $30 \times 30$ grid.}
  
  \label{fig:fig12}
\end{figure}

\begin{figure}[ht!]
\begin{minipage}{\textwidth}
\centering
\renewcommand{\tabcolsep}{0.08cm}
\begin{tabular}{ccccc}

\includegraphics[width=2.5in]{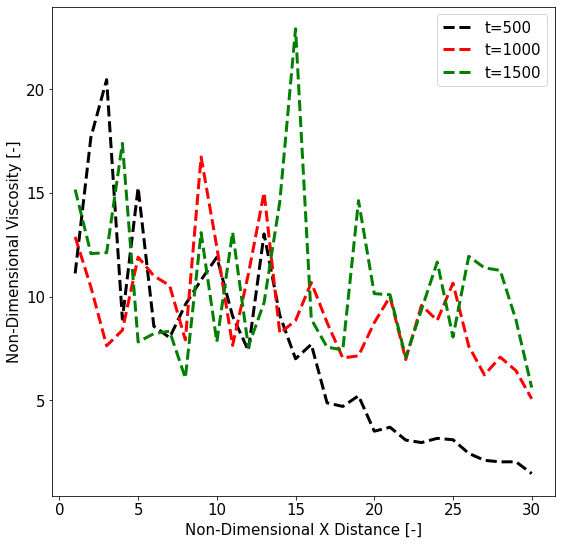}&
\includegraphics[width=2.5in]{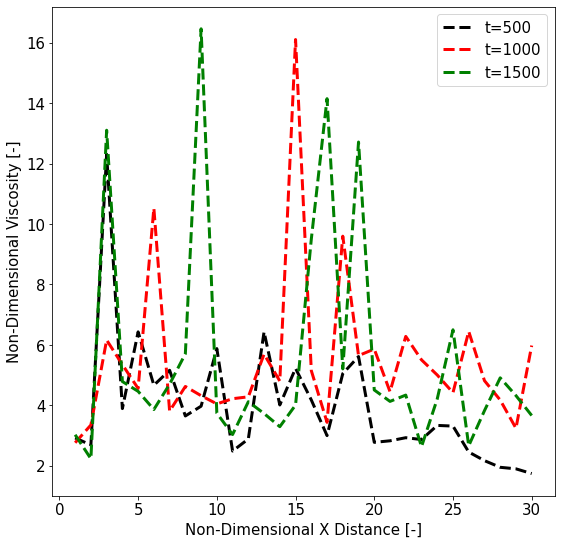}\\
\footnotesize{$\mathbf{(a)}$}&\footnotesize{$\mathbf{(b)}$}\\

\end{tabular}
\medskip
\caption{Viscosity profiles at three different time levels along the horizontal mid-section for polymers (a) Xanthane and (b) Schizophyllan.} 
\label{fig:fig4}
\end{minipage}
\end{figure}

Simulations were carried out in a quarter five spot geometry 
with heterogeneity shown in Fig.~\ref{fig:fig12}. Fig.~\ref{fig:fig4}(a) and Fig.~\ref{fig:fig4}(b) show plots of viscosity profile at the horizontal mid-section at three different times for the two polymers. We see in these figures several localized travelling waves (peaks) in viscosity at each time level for both the polymers. Development of these patterns appear to be a consequence of shear thinning property of the displacing fluids since we notice in these two figures that these patterns and their speeds are different for these two polymers even though simulations for both the polymers were carried with the same injected polymer concentration (IPC) and same injection rate (IR). These patterns and trends also change as we change the IPC and IR which will be discussed in the future when we have mathematically analyzed this complex model and have a better quantitative understanding of the nonlinear complex dynamics hidden in this model. 

\subsection{\label{sec:competing-effect}Competing effects: viscosity ratio vs shear thinning}

\begin{figure}[ht!]
 \centering
 
  \includegraphics[width=8cm, height=7cm]{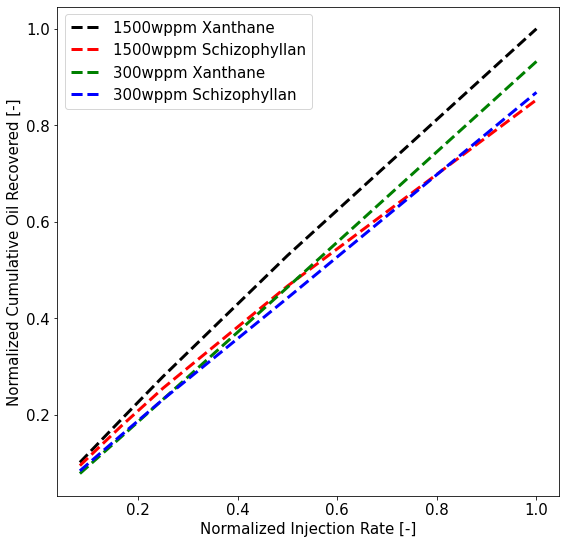}
  \caption{Cumulative oil recovered in heterogeneous rectilinear polymer simulations of Xanthane and Schizophyllan for different injected polymer concentration (IPC) (300wppm and 1500wppm) and normalized injection rate (IR) (IR=0.084, 0.167, 0.250, 0.500, 1.00). at t=100}
  
  \label{fig:fig5}
\end{figure}

Shear thinning polymers during its passage through the pores undergo shear stresses which lead to a reduction in viscosity. This viscosity reduction is proportional to the local velocity gradients which relates to the injection rates. So, if the polymer were to behave as a Newtonian fluid, high injection rates would mean high oil recovery. But due to the shear effect, a high injection rate may not necessarily correspond to high oil recovery. Therefore, modelling polymer as a non-Newtonian fluid gives a very practical benefit of assessing different injection rates and testing an ideal injection rate which helps in maintaining the viscosity ratio while avoiding the viscosity reduction due to shear thinning. This competing effect can be seen in Fig.~\ref{fig:fig5} where the injection rates were varied and cumulative oil recovery (COR) recorded for time step t=100.  

It can be seen in the figure that at lower injection rates Xanthane has slightly higher COR as compared to Schizophyllan at high IPC. But at these low injection rates, Schizophyllan has a higher COR at low IPC. As we increase the injection rates, both the polymers are subject to higher shear rates. Due to this, the advantage they provide by maintaining the viscosity ratio will start to diminish depending on the power law parameters of the polymer. We see that reduction in viscosity is higher for Schizophyllan because of diminishing returns in COR with increasing injection rate. For Xanthane, on the other hand, reduction in viscosity is not at the same rate as for Schizophyllan. Another interesting observation is flipping trend in COR for Schizophyllan. At low IR, higher IPC results in higher COR but at high IR the trend is flipped and lower IPC results in higher COR. This non-linearity is a clear indication of the two competing effects of mobility ratio and viscosity reduction due to shear thinning. 

\begin{figure}[ht!]
 \centering
 
  \includegraphics[width=16cm, height=10cm]{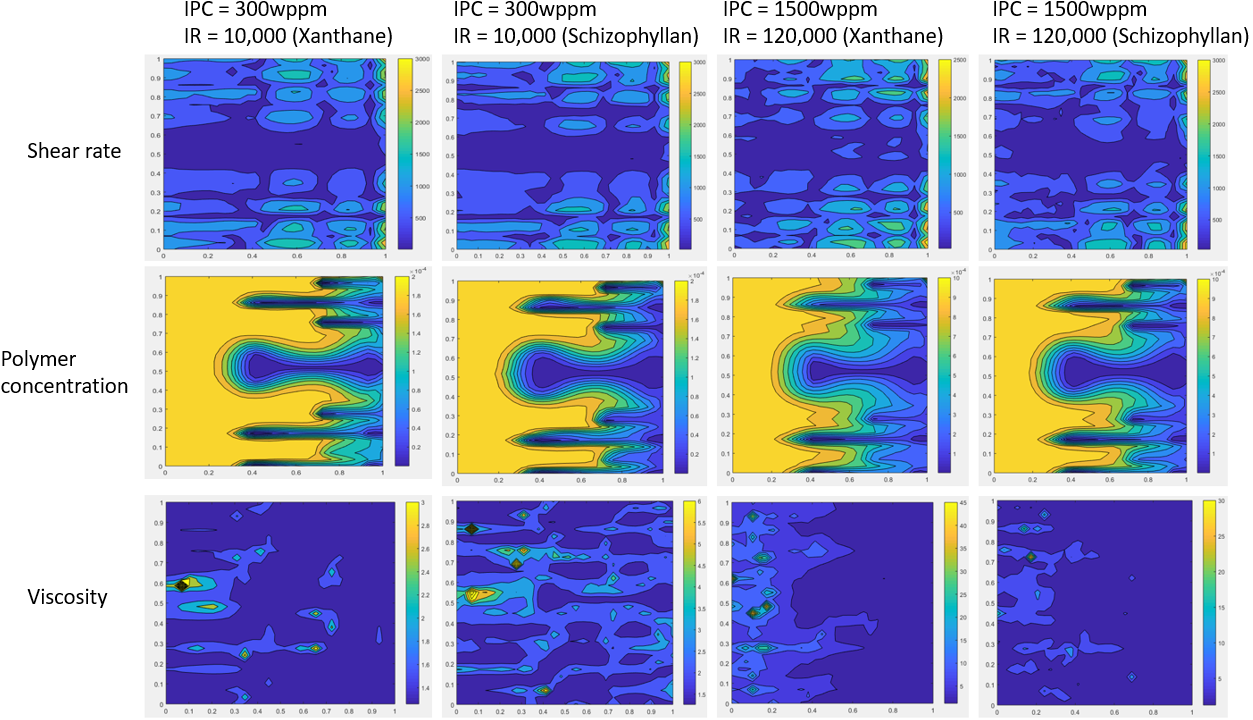}
  \caption{Contour plots of shear rate (top), polymer concentration (middle), viscosity (bottom) for rectilinear heterogeneous polymer flood simulations comparing different injection conditions and polymers at t=100}
  
  \label{fig:fig6}
\end{figure}

This competitive behavior is also observed in  Fig.~\ref{fig:fig6} for simulations in rectilinear geometry with heterogeneity generated using the following permeability tensor $\bK(\bx)$ taken from \cite{FPR2015} and shown in Fig.~\ref{fig:fig9}. 
\begin{align}\label{eq:fpr2015}
\bK(\bx) = 50 \left[ 0.5(1-10^{-7})( \sin{(6 \pi \cos{x})} \cos{(4 \pi \sin{(3y)})} -1 ) +1 \right]{\mathit {1}}, 
\end{align}
where ${\mathit {1}}$ is the second order identity tensor. The shear rate is higher in regions of high permeability. This directly relates to the ease with which the flow moves in these regions. Due to high velocity gradient, the flow experiences higher shear which eventually affects the viscosity of polysolution. Even when the concentration is same, viscosity varies for the four cases due to differences in the shear. Therefore, it can be safely concluded that while high concentration leads to higher viscosity, it is the permeability field that affects the flow movement through shear. Therefore, a shear thinning fluid will perform differently in different permeability fields. 

\subsection{Simulation in rectilinear geometry}
\subsubsection{homogeneous porous media}\label{section:simulation-in-homo}
\begin{figure}[ht!]
 \centering
 
  \includegraphics[scale=0.5]{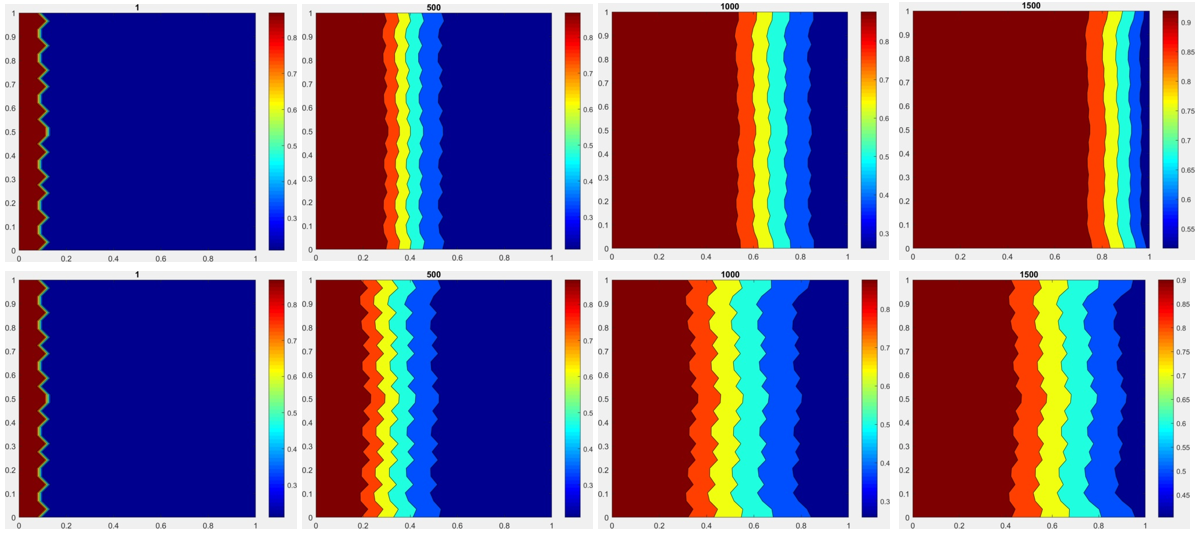}
  \caption{Temporal evolution of saturation with polymers Xanthane (top) and Schizophyllan (bottom) at IR=120000 and IPC=1500wppm in rectilinear polymer flood simulation with homogeneous permeability field.}
  
  \label{fig:fig7}
\end{figure}

First, simulations in rectilinear geometry are carried out for a homogeneous permeability field with polymers Xanthane and Schizophyllan in separate experiments. Fig.~\ref{fig:fig7} shows the saturation fields obtained with Xanthane and Schizophyllan at four time levels. This shows comparative development of the level sets of saturation due to the effect of two polymers that respond differently to the shear in the field at flooding conditions IR=120,000 and IPC=1500 wppm. It shows that the flow with polymer Xanthane moves faster than polymer Schizophyllan in this case as IR and IPC are set to their respective highest values in the range of values tested. At such low concentration of polymer Xanthane, power law index $n$ is closer to 1 and hence the viscosity has very little shear rate dependency (see \eqref{eq:powerlaw-model}). It behaves almost like a non-uniform Newtonian fluid, non-uniform embodies the fact that aqueous phase viscosity will still vary in space and time due to the data and time dependency of  $\rho$ and $\varepsilon$ which the power law~\eqref{eq:powerlaw-model} depends on. On the other hand, for Schizophyllan viscosity depends on shear rate with the index $n$ around 0.7 and affects the viscosity profoundly in comparison to Xanthane. This difference justifies the difference in the speed with which the saturation fronts move in these two cases. Perhaps polymer Xanthane undergoes significant shear thinning in comparison to Schizophyllan and hence moves faster due to reduction in viscosity.

Fig.~\ref{fig:fig7} also shows narrow mixing regions of mild viscous fingers for both the polymers, though the mild fingers in the case of Schizophyllan are somewhat more pronounced. These fingers are mild in comparison to classical ones for displacements involving Newtonian flows because mobility ratio across the interfaces in this figure is position dependent and can change from favorable to unfavorable constantly along an interface because of data dependent shear thinning property of the fluids as discussed earlier. Finger widths in these cases have been computed using the procedure described in appendix~\ref{appendix:fingerwidth-calculation}. Fig.~\ref{fig:fig8}(a) shows finger width growth for polymers Xanthane and Schizophyllan at IR=10,000 and IPC=300 wppm. Fig.~\ref{fig:fig8}(b) shows the effect of finger width growth on cumulative oil recovery (COR). It is evident that Schizophyllan with higher finger width shows higher recovery as compared to Xanthane. However, this trend is reversed for higher values of IPC and IR. Plots in Fig.~\ref{fig:fig8}(c),(d) show Xanthane to have higher finger width and COR at IR=120,000 and IPC=1500 wppm. This clearly shows the importance of choosing a polymer based on the IR and IPC conditions in a given polymer flood.

\begin{figure}[ht!]
\begin{minipage}{\textwidth}
\centering
\renewcommand{\tabcolsep}{0.08cm}
\begin{tabular}{ccccc}

\includegraphics[width=2.0in]{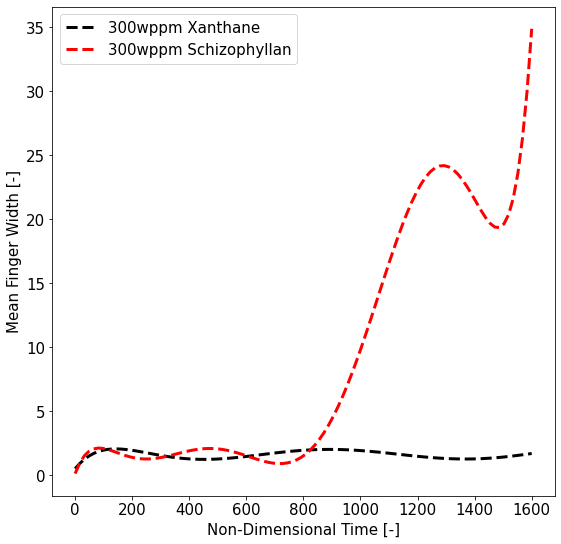}&
\includegraphics[width=2.0in]{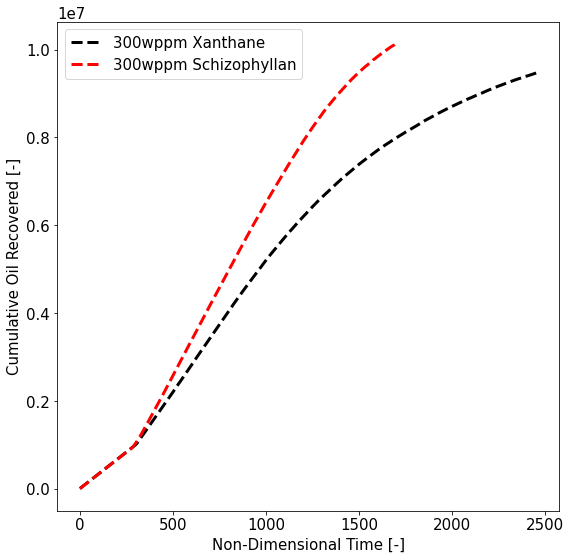}\\
\footnotesize{$\mathbf{(a)}$}&\footnotesize{$\mathbf{(b)}$}\\
\includegraphics[width=2.0in]{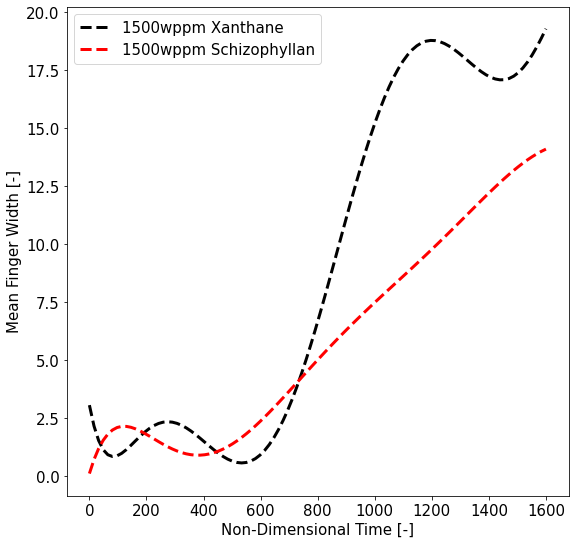}&
\includegraphics[width=2.0in]{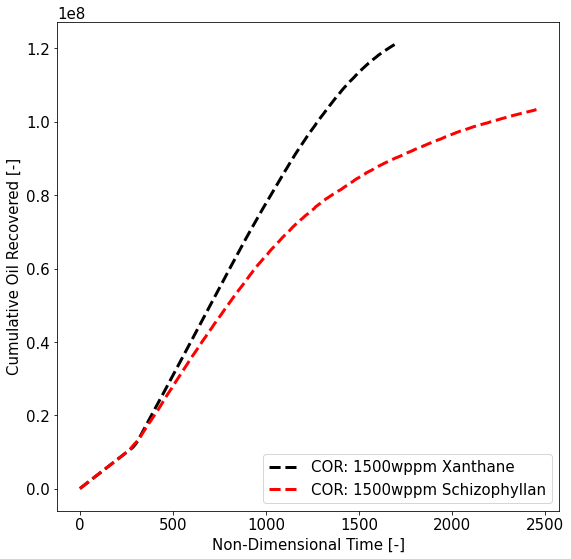}\\
\footnotesize{$\mathbf{(c)}$}&\footnotesize{$\mathbf{(d)}$}\\
\end{tabular}
\medskip
\caption{COR and MFW from rectilinear homogeneous polymer flood simulation. Top row: Comparison of Xanthane and Schizohyllan at IR=10,000 and IPC=300wppm (a) Mean Finger Width (MFW) and (b) Cumulative Oil Recovery (COR).
Bottom row: Comparison of Xanthane and Schizohyllan at IR=120,000 and IPC=1500wppm (c) Mean Finger Width (MFW) and (d) COR} 
\label{fig:fig8}
\end{minipage}
\end{figure}


\begin{figure}[ht!]
 \centering
 
  \includegraphics[scale=0.3]{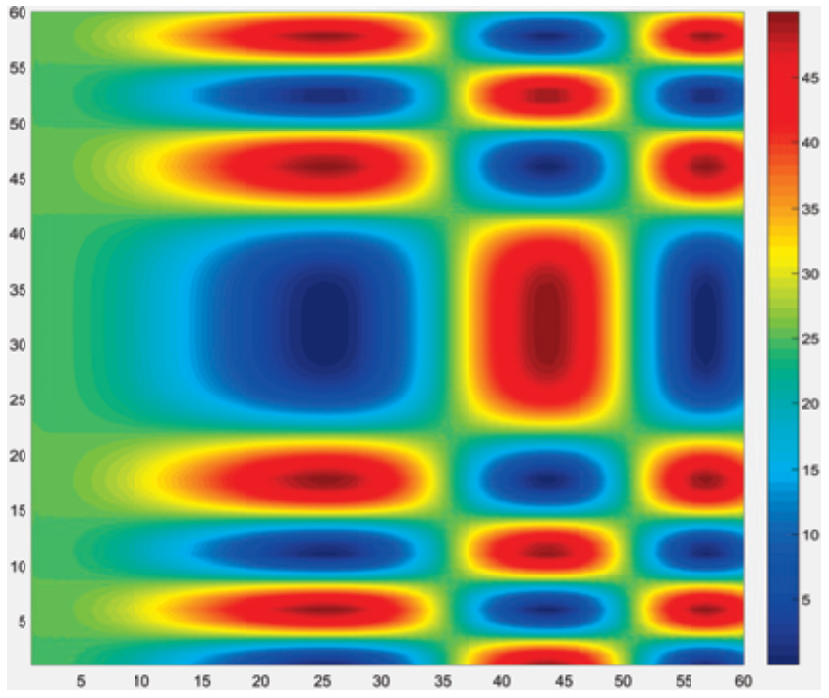}
  \caption{The heterogeneous permeability field given by \eqref{eq:fpr2015} and taken from \cite{FPR2015} is plotted in a rectilinear geometry with a $60 \times 60$ spatial resolution. Red regions represent higher permeability while the blue regions represent lower permeability.}
  
  \label{fig:fig9}
\end{figure}
\subsubsection{heterogeneous porous media}
\begin{figure}[ht!]
 \centering
 
  \includegraphics[scale=0.4]{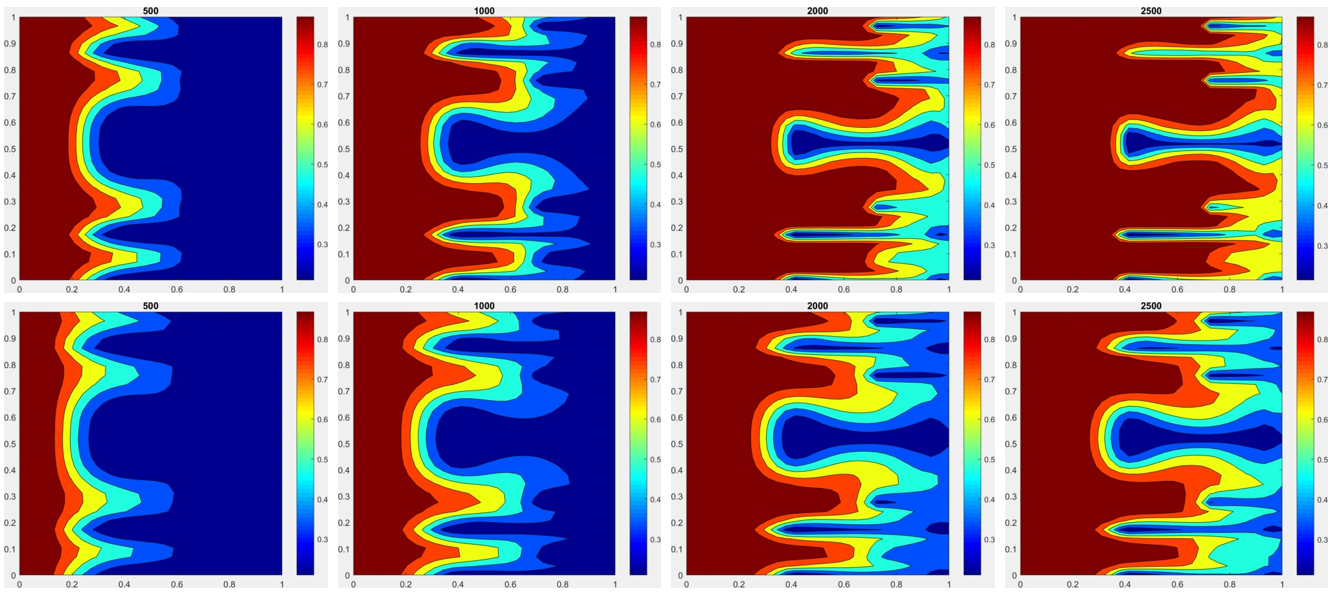}
  \caption{Temporal evolution of saturation with polymers Xanthane (top) and Schizophyllan (bottom) at IR=120000 and IPC=1500 wppm in a rectilinear geometry with heterogeneous permeability field shown in Fig.\ref{fig:fig9}.}
  
  \label{fig:fig10}
\end{figure}

\begin{figure}[ht!]
\begin{minipage}{\textwidth}
\centering
\renewcommand{\tabcolsep}{0.08cm}
\begin{tabular}{ccccc}

\includegraphics[width=2.5in]{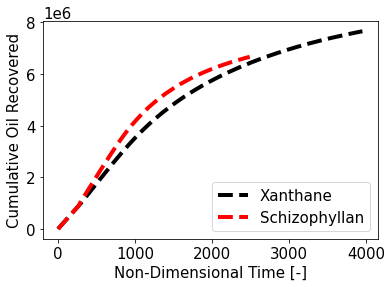}&
\includegraphics[width=2.5in]{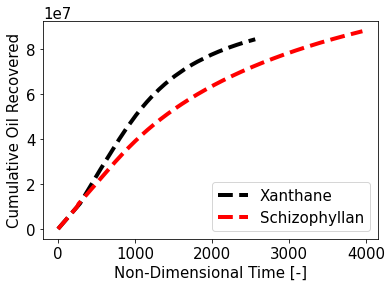}\\
\footnotesize{$\mathbf{(a)}$}&\footnotesize{$\mathbf{(b)}$}\\

\end{tabular}
\medskip
\caption{(a) Comparison of COR for Xanthane and Schizohyllan at IR=10,000 and IPC=300wppm (a) and (b) Comparison of COR for Xanthane and Schizohyllan at IR=120,000 and IPC=1500wppm in rectilinear heterogeneous polymer flood simulation.} 
 
\label{fig:fig11}
\end{minipage}
\end{figure}

Next, heterogeneous permeability field shown in Fig.~\ref{fig:fig9} is used for simulation in core flood. 
The saturation fields and COR are shown in Fig.~\ref{fig:fig10} and Fig.~\ref{fig:fig11} respectively. Although most of the flooding is now affected by this heterogeneous permeability field, but for the same IR and IPC clear differences can be noticed on the predicted COR for Schizophyllan and Xanthane. This is a clear indicator that choice of polymer also depends on the permeability field. For the given case, Xanthane seems to outperform Schizophyllan in terms of COR. But given a different set of IR, IPC and permeability field, the choice of polymer may change. In practice, therefore permeability fields also influence the choice of polymer and other outcomes from shear thinning polymer flooding. The fingering nature of the flow pattern seen in Fig.~\ref{fig:fig10} is due to channeling effect as there are regions of high permeability.

\subsection{Quarter five spot simulation}
\begin{figure}[ht!]
 \centering
 
  \includegraphics[scale=0.4]{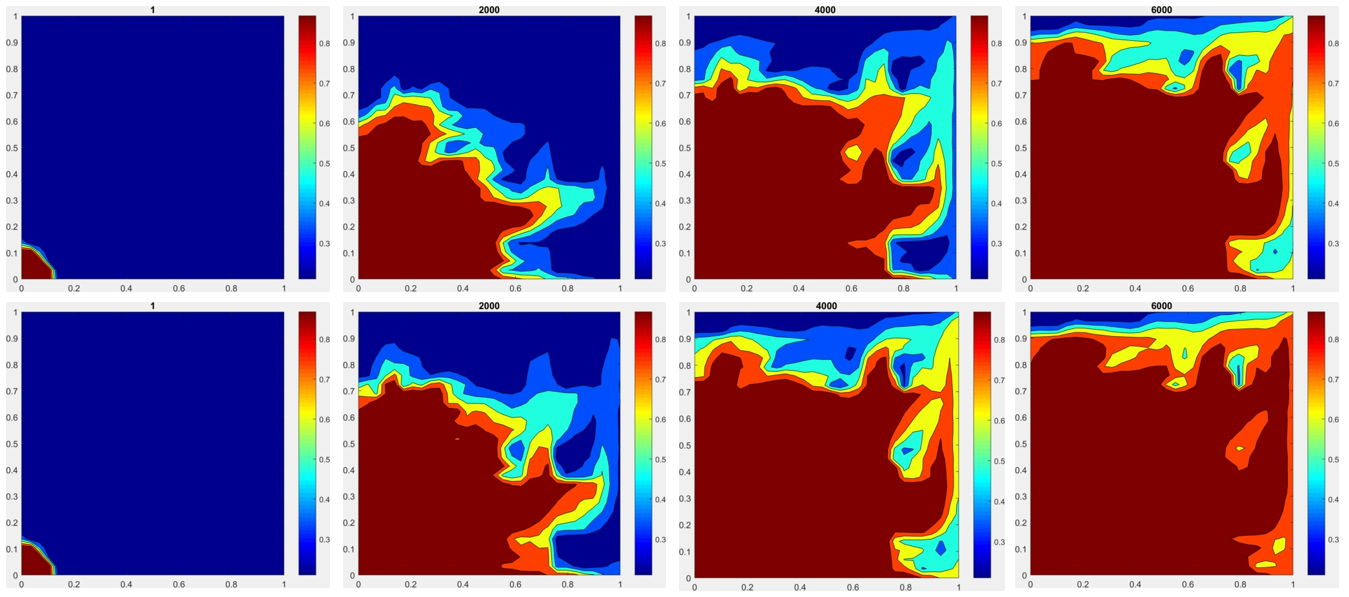}
  \caption{Temporal evolution of saturation for Xanthane (top) and Schizophyllan (bottom) at IR=120000 and IPC=1500 wppm in quarter-five spot heterogeneous polymer flood simulation}
  
  \label{fig:fig13}
\end{figure}

\begin{figure}[ht!]
\begin{minipage}{\textwidth}
\centering
\renewcommand{\tabcolsep}{0.08cm}
\begin{tabular}{ccccc}

\includegraphics[width=2.5in]{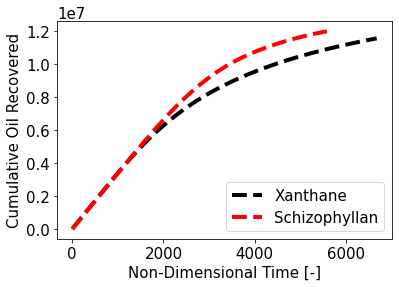}&
\includegraphics[width=2.5in]{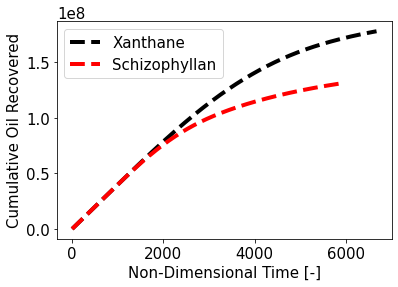}\\
\footnotesize{$\mathbf{(a)}$}&\footnotesize{$\mathbf{(b)}$}\\

\end{tabular}
\medskip
\caption{Comparison of cumulative oil recovered in quarter five-spot heterogeneous polymer flood simulation with polymers Xanthane and Schizohyllan at (a) IR=10,000 and IPC=300 wppm and at (b) IR=120,000 and IPC=1500 wppm.} 
\label{fig:fig14}
\end{minipage}
\end{figure}

Finally, a quarter five spot geometry with a heterogeneous logarithmic permeability field shown in Fig.~\ref{fig:fig12} is used for simulation. Fig.~\ref{fig:fig13} and Fig.~\ref{fig:fig14} show the saturation field and COR for two different polymers. Here again the influence of the permeability field is evident. Also the comparative simulations show significant difference as the variable permeability field leads to higher shear and subsequent variation for different polymers used in the flooding. In simulations with heterogeneous permeability fields, mean finger width does not really help in understanding the underlying physics as the flow becomes very complex due to the interaction of heterogeneous permeability, time dependent variable viscosity field and travelling viscosity waves generated by the shear thinning effect which has been alluded to earlier. The only metric that can help understand the flow then becomes the COR shown in  Fig.~\ref{fig:fig14} for both the polymers.

 
  

\subsection{Effect of injection rate (IR) and injected polymer concentration (IPC) on Cumulative Oil Recovered (COR)}

\begin{figure}[ht!]
 \centering
 
  \includegraphics[scale=0.4]{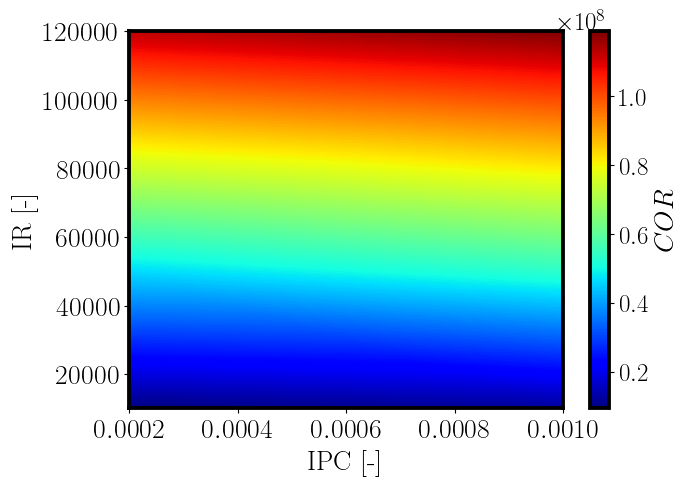}
  \caption{COR at water breakthrough for Xanthane}
  \label{fig:fig15}
\end{figure}

\begin{figure}[ht!]
 \centering
 
  \includegraphics[scale=0.4]{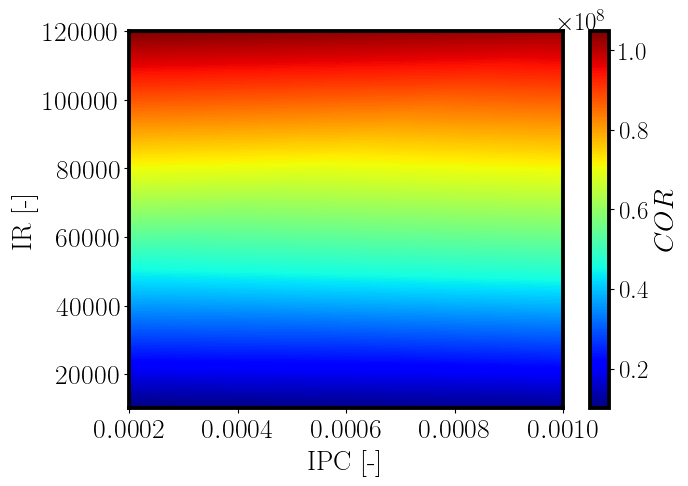}
  \caption{COR at water breakthrough for Schizophyllan}
  \label{fig:fig16}
\end{figure}

The effect of IR and IPC is polymer specific and can be quantified on a common parameter namely COR. Figures \ref{fig:fig15} and \ref{fig:fig16} show the COR for varying IR and IPC conditions for homogeneous rectilinear polymer flooding simulation. It can be seen that both the polymers predict COR with a strong dependence on IR. However, IPC seems to be a less important parameter and the degree to which it influences the COR differs for the two polymers. Xanthane has a stronger dependence on IPC for predicting the COR as compared to Schizophyllan. This information enables correct estimation of polymer required for any flooding process and also indicates the importance of IR that will be required for optimum recovery. 

\subsection{Sensitivity of the power law parameters ($\varepsilon$ and $\eta$)}
Significance and implications of the way the power law viscosity model~\eqref{eq:powerlaw-model} is implemented here are exemplified in this section. This model is not an ordinary power law model with fixed parameters, rather values of the parameters $n$ and $\varepsilon$ are based on the type and concentration $c(x,t)$ of polymer. Since the concentration of polymer is evolving in space and time according to its own transport equation and the power law viscosity model coefficients at every point in space and time are derived from experimental data, effectively this means the classical shear thinning fluid, if thought to have constant values of the power law parameters, itself is changing in space and time. In this way we treat the flow in the most physical way. To the best of the authors' knowledge, to-date this type of space and time dependent viscosity model has not been implemented in a full scale shear thinning polymer flood simulations.

\begin{figure}[ht!]
 \centering
 
  \includegraphics[width=8cm, height=7cm]{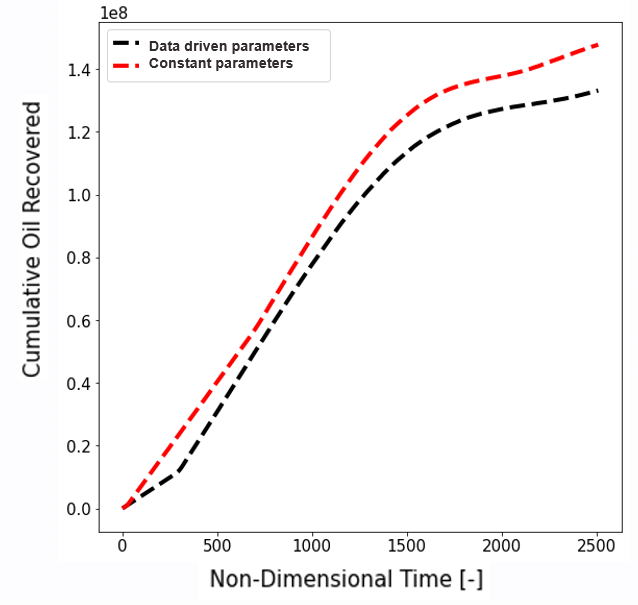}
  \caption{ Plots of cumulative oil recovered versus time for constant values of shear thinning parameters and for data driven values of space-time varying parameters in rectilinear homogeneous polymer flooding for IR=120000 and IPC=1500 wppm.}
  
  \label{fig:fig17}
\end{figure}

\begin{figure}[ht!]
 \centering
 
  \includegraphics[scale=0.4]{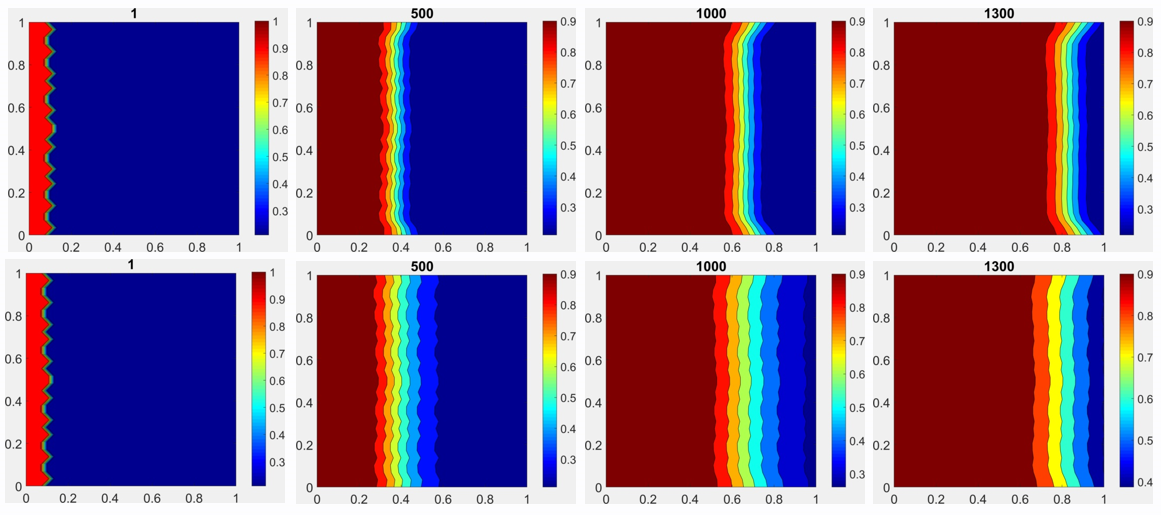}
  \caption{Temporal evolution of saturation in polymer flood simulation with constant power law coefficients (top) and variable power law coefficients (bottom) in rectilinear homogeneous polymer flooding at IR=120000, IPC=1500wppm.}
  \label{fig:fig18}
\end{figure}

Fig.~\ref{fig:fig17} shows plots of cumulative oil recovery (COR) versus time for constant values of $n$ and $\varepsilon$ (average values over the entire range chosen) and also for experimental data driven values of these two parameters. Interestingly, difference is observed right from the beginning and is very significant at breakthrough. The graph has three distinct regions. The first region from t=0 to t=300. The second one from t=300 to t=1600 and the third one from t=1600 to t=2500. We see in this figure that simulation with experimental data dependent parameters results in significant variation in COR rate over all three regions as compared to a more or less constant COR rate for constant parameter simulation. In the first region, COR rate is lower for the data dependent parameters than that for constant values of parameters indicating that higher effective polymer viscosity makes the flow slow. This is also seen in Fig.~\ref{fig:fig18} where level sets of saturation at four different time levels with constant power law coefficients (top) and with variable power law coefficients (bottom) in rectilinear homogeneous polymer flooding are compared. The parameters chosen are for high polymer concentration relating to high viscosity. In the second region, due to now mobilized flow which corresponds to higher shear rates, the viscosity starts to decrease even if the concentration stays the same leading to higher COR rate. Finally, in the third region the COR rate decreases due to reduction in sweep efficiency as some aqueous phase reaches the production well. In this region, both constant and data dependent parameter simulations show similar trend. The maximum variation at water breakthrough is around 33$\%$. This is a clear indication of the importance of data dependent shear thinning polymer flood simulations. It is worth observing in Fig.~\ref{fig:fig18} that saturation in the data dependent parameter case is more diffused than in the constant parameter case. This can be attributed to the viscosity which is not only varying in space but is also dependent on the local concentration of polymer and therefore smearing the so-called interface. 


\usetikzlibrary{%
	decorations.pathreplacing,%
	decorations.pathmorphing%
}

\subsection{Sensitivity to ${\rm K}_{max}$}
\begin{figure}[ht!]
 \centering
 
  \includegraphics[scale=0.5]{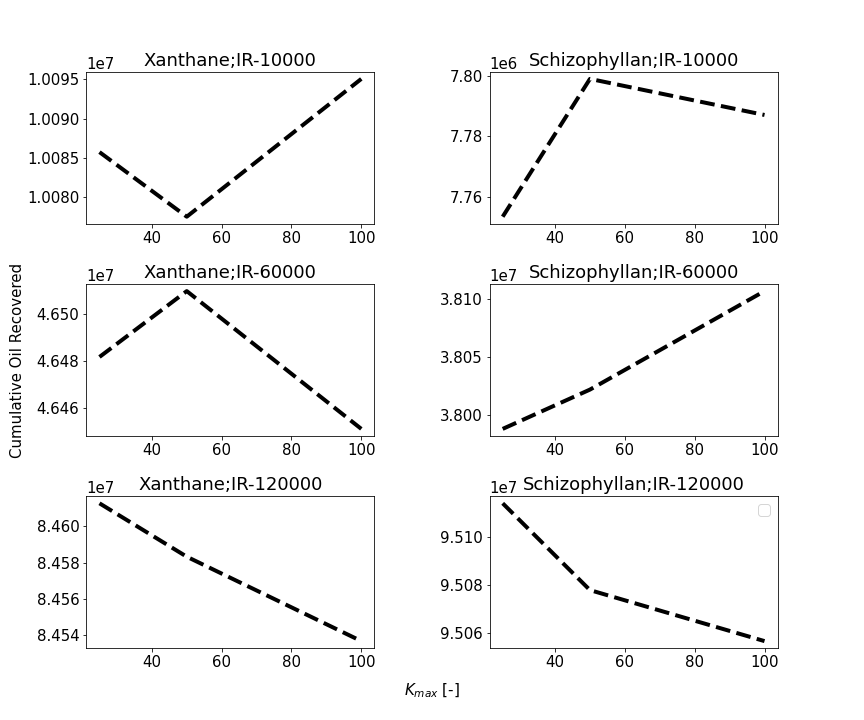}
  \caption{Cumulative Oil Recovered (at water breakthrough) showing sensitivity to  ${\rm K}_{max}$ for different injection rates and polymers (Comparison made for heterogeneous rectilinear geometry)}
  
  \label{fig:fig19}
\end{figure}

Any given polymer flooding simulation is highly sensitive to the permeability field. In this section, effect of varying the maximum value, ${\rm K}_{max}$, of permeability has been discussed. Multiple simulations with the two polymers at different injection rates and ${\rm K}_{max}$ values were run. Fig.~\ref{fig:fig19} shows the cumulative oil recovered at water breakthrough for different conditions. A clear trend is not seen which is expected given the highly non-linear and data driven nature of the problem. However there are some trends that are seen across the two polymers. For lower injection rates, increasing the ${\rm K}_{max}$ value is seen to increase the COR. At some particular cutoff this trend flips which is expected to be lower than IR=60000 for Xanthane but higher for Schizophyllan. As seen from the IR=60000 the overall effect of increasing ${\rm K}_{max}$ is different for Xanthane and Schizophyllan, however the trend is decreasing for both the polymers at the maximum injection rate (IR=120000). So to generalize, increasing the value of ${\rm K}_{max}$ increases (decreases) COR for lower (higher) injection rates. This can be explained from the relation of shear rate with viscosity for complex fluids. At lower injection rates the flow experiences lower shear rate resulting in higher viscosity of the shear thinning fluid. Effect of this higher viscosity is opposed by the lower resistance to the flow from the medium at higher ${\rm K}_{max}$ value. It appears that this second effect wins over the first one with the net effect of increasing overall COR at water breakthrough. 

\begin{figure}[ht!]
 \centering
 
  \includegraphics[scale=0.5]{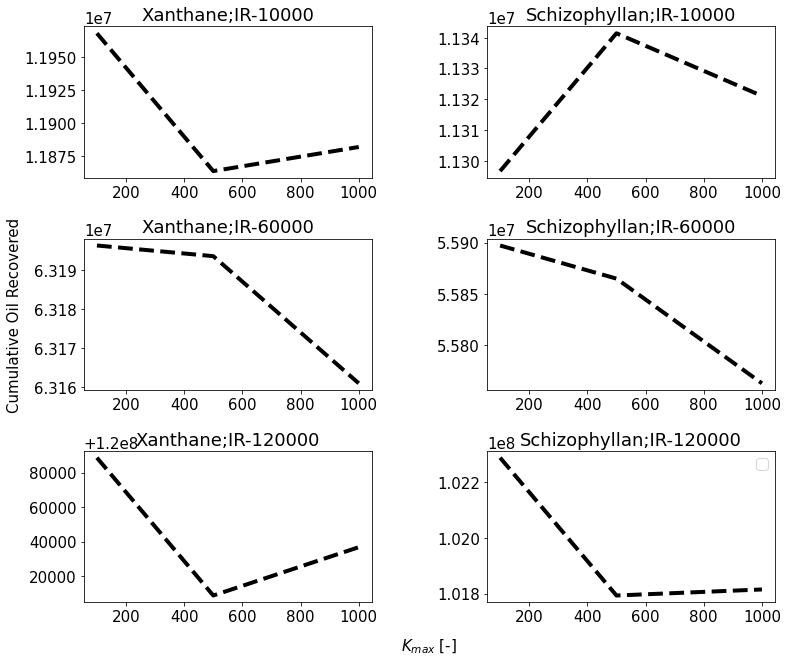}
  \caption{Cumulative Oil Recovered (at water breakthrough) showing sensitivity to  ${\rm K}_{max}$ for different injection rates and polymers (Comparison made for homogeneous rectilinear geometry)}
  
  \label{fig:fig20}
\end{figure}
A similar trend is seen for the homogeneous case. Fig.~\ref{fig:fig18} shows the cumulative oil recovered for different polymers and injection rate across  ${\rm K}_{max}$ values. However, in this case for the chosen range of  ${\rm K}_{max}$ values, the cutoff value is not seen for Xanthane but is seen to be between IR=60,0000 and IR=120,000 for Schizophyllan. 

\section{\label{sec:conslusions}Conclusions}

A dynamic empiric coefficient based shear thinning model of polymer flooding has been implemented in an in-house code for modelling multi-component multi-phase fluid flow in porous media. Simulations with this data driven model using this code have been performed for two shear thinning polymers, Xanthane and Schizophyllan, at various values of the parameters of the problem. Numerical results which are qualitatively consistent with physics show the merits of this easy-to-implement inexpensive and fast method. Time dependent viscosity profiles in the flow field show a train of travelling viscosity waves which are not yet fully understood theoretically and opens the door for future research for a better theoretical understanding of the complex PDE model for the shear tinning polymer flooding. The simulations also show (i) competing effects of shear thinning and mobility ratio; (ii) injection conditions such as injection rate and injected polymer concentration influence the choice of polymers to optimise cumulative oil recovery (these effects are relatively much more significant in the case of Schizophyllan compared to Xanthane); (iii) permeability field also affects the choice of polymer as polymers show varying movement for different shear rates that are caused by heterogeneity; and (iv) dynamically evolving travelling wave patterns of viscosity profiles which interact with the underlying flow and the heterogeneity field to generate narrow region of fingers which need to be further probed in the future. The significance of this work is that it shows a simple way, yet accurate enough, to incorporate shear thinning effect of polymer in an otherwise Newtonian model of multiphase multicomponent porous media flow model and provides an effective yet easy approach to make design choices of polymers for CEOR in any given flooding condition. We summarize below some of the specific results.
\begin{enumerate}
    \item Schizophyllan polymer viscosity decreases faster with increasing shear rate as compared to Xanthane.
    \item Xanthane performs better in terms of COR for higher IPC and IR conditions while Schizophyllan shows higher COR for lower IPC and IR. This fact is reflected from the MFW which in turn shows that the mean finger width is higher for the polymer producing higher COR for a given condition.
    \item Competing effects of viscosity ratio and shear thinning behavior is seen in Schizophyllan polymer for the range of IR and IPC tested. COR is higher for higher IPC at lower injection rates which reduces to a COR less than the case of lower IPC at higher injection rates. This indicates that lower IPC might be preferable for some polymers at higher injection rates.
    \item Permeability field directly affects the shear rates and therefore the viscosity of the aqueous solution. Polymers might perform differently in different permeability fields when shear thinning behavior is accounted for.
    \item The percentage difference for COR is at $33\%$  for shear thinning model when comparing varying parameters and constant power law parameters. This shows the importance of these models to predict the overall recovery in a given polymer flood simulation. These percentages may vary depending on the IPC, IR and permeability.
\end{enumerate}

\appendix
\section{Appendix}\label{sec:appendix}
\subsection{Algorithm}\label{appendix:algorithm}
Here we outline the algorithm for the SP-flood simulation that includes shear thinning effect of polymer. The algorithm for the polymer flood is essentially a special case of the same with zero concentration for the surfactant. The step-by-step algorithm built on the one described in Daripa \& Dutta~\cite{daripa2017modeling} is given below. To accomplish some of the steps below, one needs to refer Daripa \& Dutta~\cite{daripa2017modeling}.
\begin{enumerate}
\item Define the Cartesian grid in the domain using equal, uniform grid sizes in both the axes. Generate the finite element mesh.
\item Generate a heterogeneity field on this grid.
\item Choose an initial interface separating the injected fluid from the resident fluid.
\item Set the model parameters: $\mu_o$, $\mu_w$, $s_{ro}^{\sigma 0}$, $s_{ra}^{\sigma 0}$.
\item Initialize the state variables $s$, $c$ and $\Gamma$ as 
\begin{align*}
s_0 = 
	\begin{cases}
    1 \hfill &  x \in \Omega^+\\
    s_0^{\sigma0} \hfill & x \in \Omega^-\\
    \end{cases}, \qquad
c_0 = 
	\begin{cases}
    0.01 \hfill &  x \in \Omega^+\\
    0 \hfill & x \in \Omega^-\\
    \end{cases}, \qquad
\Gamma_0 = 
	\begin{cases}
    0.005 \hfill &  x \in \Omega^+\\
    0 \hfill & x \in \Omega^-\\
    \end{cases}.    
\end{align*}
\item  Calculate $\sigma(s^n,\Gamma^n)$, $\mu_a(c^n,\nabla \bv^{n-1})$, $s_{ra}(s^n,\Gamma^n)$,$s_{ro}(s^n,\Gamma^n)$, $\lb_a(s^n,c^n,\Gamma^n)$, $\lb_o(s^n,c^n,\Gamma^n)$, $\lb(s^n,c^n,\Gamma^n)$ using $s^n$, $c^n$, $\Gamma^n$ which are values of $s$, $c$ and $\Gamma$ respectively at the $n^{th}$ time level. \label{5}
\item Solve the global pressure equation to get $p^{n}$ and subsequently compute $\bv^{n}$.\label{6}
\item Use $\bv^{n}$, $s^n$, $c^n$, $\Gamma^n$ and the quantities calculated in Step \ref{5}, to solve for $s^{n+1}$, $c^{n+1}$ and $\Gamma^{n+1}$, thus completing a full time step. \label{7}
\item If breakthrough is achieved: then stop; else update $n=n+1$ and repeat from Step \ref{5} above.
\end{enumerate}
In order to reduce computational cost, a few iterations of Steps \ref{5} and \ref{7} are done before updating the pressure in Step \ref{6}. The pseudocode (see Algorithm \ref{eor}) and flow-chart (see Fig.~\ref{fig:flow-chart}) for the procedure are given here.

\begin{algorithm}[H]
\SetAlgoLined
\caption{SP flooding simulation}
\label{eor}
\DontPrintSemicolon \tcc*[l]{\small Set up Cartesian grid, FE Mesh, permeability field and model parameters}
\vskip 1ex \normalsize
Set $i,j = 0, \ldots , M; \; \bx_{ij} = \left(\dfrac{i}{M},\dfrac{j}{M}\right);$  \tcc*{\small($M\times M$ \textit{is the grid size})} \normalsize
\vskip.5ex
Set $\Sigma = $ \textit{Initial interface}; \tcc*{\small $\Sigma = \dl \Omega^+ \cap \dl \Omega^-$}\normalsize
\vskip .5ex
Set $\bK(\bx)$ 
\vskip .5ex
Set $\mu_o, \, \mu_w, \, s_{ro}^{\sigma 0}, \, s_{ra}^{\sigma 0}, \, \tilde{q} = \textit{values from \ref{table:1}}$; \;     %
\vskip 1.5ex
\DontPrintSemicolon \tcc*[l]{\small Initialization}
\vskip 1.5ex \normalsize
Set $t = 0;  \;  \Delta t = \dfrac{1}{N}; \; Tstop = N\Delta t $; \tcc*{\small $N$ chosen for accuracy} \vskip .5ex
\For{i = 0, \ldots , M}{ \For {j = 0, \ldots , M}{
Set $(s, \,c, \, \Gamma)(\bx_{ij},0)= \begin{cases}
(1, 0 , 0) \hfill & \bx_{ij} \in \Omega^+\\
(s_0^{\sigma0}, c_0, \Gamma_0) \hfill & \bx_{ij} \in \Omega^-\\
\end{cases}$ ; \;}}
\vskip 1.5ex
\DontPrintSemicolon \tcc*[l]{\small Computation loop}
\vskip 1.5ex \normalsize
\While{$\left(s(\bx_{M,M},t) \leq 1-s_0^{\sigma0} \;\; \&\& \;\; t <Tstop \right)$} {
Compute $\{\sigma, \, \mu_a, \,s_{ra}, \,s_{ro},\, \lb_a, \, \lb_o, \, \lb, \, p_c\}\; \text{using} \;(s^n, c^n, \Gamma^n, \bv^{n-1})$; \;
\vskip .5ex
Use data driven power law model to calculate $\mu_a$; \; 
\vskip .5ex
Update $\lambda_a, \lambda, f_a, D(s,c,\Gamma)$ using $\mu_{a}$; \;
\vskip .5ex

Solve global pressure equation for $p^{n}, \bv^n $; \;
\vskip .5ex
Recompute $\{s_{ra}, \, s_{ro}, \, \lb_a, \, \lb_o, \, \lb\} \; \text{using} \; (s^n, c^n, \Gamma^n, \bv^{n})$; \;
\vskip .5ex
Solve transport equations for $s^{n+1}$, $c^{n+1}$ and $\Gamma^{n+1}$; \;
\vskip .5ex
Set $t = t+ \Delta t$; \;
}
\end{algorithm}

\tikzstyle{decision} = [diamond, draw, fill=green!30,
    text width=4.5em, text badly centered, node distance=3cm, inner sep=0pt]
\tikzstyle{process} = [rectangle, draw, fill=orange!30,
    text width=4em, text centered, rounded corners, minimum height=4em]
    \tikzstyle{input} = [rectangle, draw, fill=blue!30,
    text width=5em, text centered, rounded corners, minimum height=4em]
\tikzstyle{line} = [draw, -latex']
\tikzstyle{start} = [draw,rectangle, rounded corners,fill=red!30, node distance=3cm,
    minimum height=3em, text width = 5em]
\tikzstyle{stop} = [draw, rectangle, rounded corners,fill=red!60, node distance=3cm,
    minimum height=2em, minimum width = 2em]

\begin{omitext}
\begin{figure}[ht!]
\begin{tikzpicture}[node distance = 2cm,auto][h]
    \node [start] (init) {Setup grid \\ Load $\bK(\bx)$ }; 
    \node [input, below of =init] (init2) {Set: $\mu_o,\mu_w$,\\$s_{ro}^{\sigma 0},s_{ra}^{\sigma 0}$};
    \node [input, below of=init2] (init3) {Initialize: \\  $s_0,c_0,\Gamma_0$};
    \node [process, right of=init3,xshift=.6cm] (interm) {Compute $\sigma, \mu_a, s_{ra},$\\$s_{ro},\lambda_a, \lambda_o$};
    \node [process, right of=interm,xshift=.4cm] (visc) {Calculate $\mu_a$  };
    \node [process, right of=visc,xshift=.4cm] (evaluate) {Calculate $u,v$};
    \node [process, above of=evaluate,node distance=2cm] (interm2) {Compute $\sigma, p_c$};
    \node [process, right of=evaluate,xshift=.4cm] (solve) {Solve for $s,c,\Gamma$};
    \node [process, below of=evaluate, node distance=3cm,xshift = 1.3cm] (update) {Update model};
    \node [decision, right of=solve] (decide) {Water breakthrough?};
    \node [stop, below of=decide, node distance=3cm] (stop) {\bf stop};
    \path [line] (init) -- (init2);
    \path [line] (init2) -- (init3);
    \path [line] (init3) -- (interm);
    \path [line] (interm) -- (visc);
    \path [line] (interm) |- (interm2);
    \path [line] (interm2) -| (solve);
    \path [line] (visc) -- (evaluate);
    \path [line] (evaluate) -- (solve);
    \path [line] (solve) -- (decide);
    \path [line] (decide) -- node [below] {no} (update);
    \path [line] (update) -| (interm);
    \path [line] (decide) -- node[right] {yes}(stop);
\end{tikzpicture}
\caption{Flow-chart for SP-flood simulation that includes shear thinning effect of polymer. For polymer-flood simulation, we set $\Gamma_o=0$ and we do not solve for $\Gamma$. {\color{red} Replace this flow chart with the original one in Daripa \& Datta}} \label{fig:flow-chart}
\end{figure}
\end{omitext}

\begin{figure}[ht!]
\begin{tikzpicture}[node distance = 2cm,auto]
    \node [start] (init) {Setup grid \\ Load $\bK(\bx)$ }; 
    \node [input, below of =init] (init2) {Set: $\mu_o,\mu_w$,\\$s_{ro}^{\sigma 0},s_{ra}^{\sigma 0}$};
    \node [input, right of=init2,xshift=.7cm] (init3) {Initialize: \\  $s_0,c_0,\Gamma_0$};
    \node [process, right of=init3,xshift=.6cm] (interm) {Compute $\sigma, \mu_a, s_{ra},$\\$s_{ro},\lambda_a, \lambda_o$};
    \node [process, right of=interm,xshift=.4cm] (evaluate) {Calculate $u,v$};
    \node [process, above of=evaluate,node distance=2cm] (interm2) {Compute $\sigma, p_c$};
    \node [process, right of=evaluate,xshift=.4cm] (solve) {Solve for $s,c,\Gamma$};
    \node [process, below of=evaluate, node distance=2cm,xshift = 1.3cm] (update) {update model};
    \node [decision, right of=solve] (decide) {Water breakthrough?};
    \node [stop, below of=decide, node distance=3cm] (stop) {\bf stop};
    \path [line] (init) -- (init2);
    \path [line] (init2) -- (init3);
    \path [line] (init3) -- (interm);
    \path [line] (interm) -- (evaluate);
    \path [line] (interm) |- (interm2);
    \path [line] (interm2) -| (solve);
    \path [line] (evaluate) -- (solve);
    \path [line] (solve) -- (decide);
    \path [line] (decide) -- node [below] {no} (update);
    \path [line] (update) -| (interm);
    \path [line] (decide) -- node[right] {yes}(stop);
\end{tikzpicture}
\caption{Flow-chart for SP flooding simulation} \label{fig:flow-chart}
\end{figure}
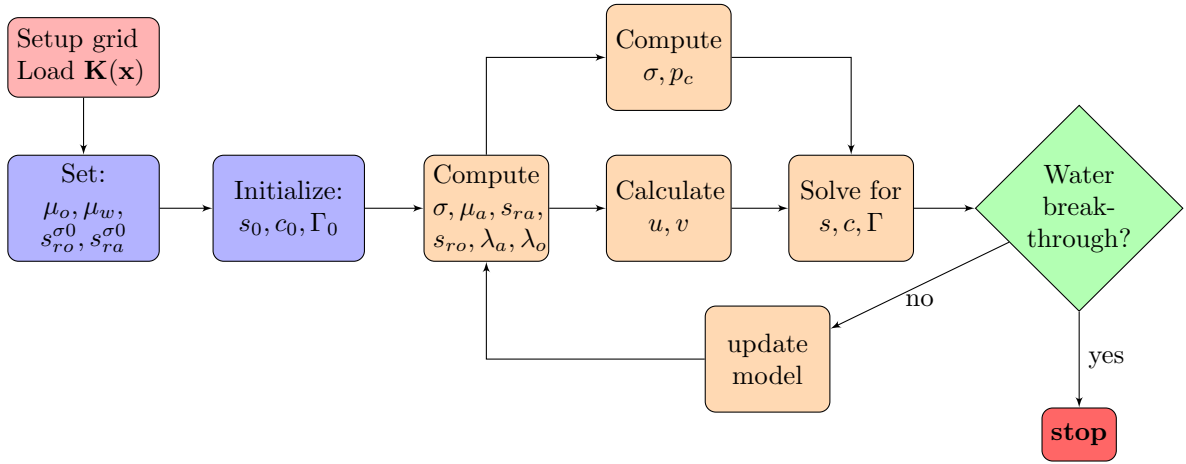

\begin{omitext}
\subsection{Proof for shear thinning model}\label{appendix:proof}

Relationship between stress tensor and deformation tensor is given by:

\begin{equation}\Pi = f({\bf\nabla}{\bv})\end{equation}

\begin{equation}{\bf \nabla}\bv = \frac{1}{2}({\bf\nabla} \bv + {\bf\nabla} {\bv}^T) + \frac{1}{2}({\bf\nabla}\bv - {\bf\nabla}{\bv}^T) = D + \Omega\end{equation}

Since $\Pi$ is frame indifferent quantity from principle of material objectivity and D being frame indifferent and $\Omega$ not being frame indifferent gives the following identity:

\begin{equation}\Pi=f(D)\end{equation}

Now, $f(D)$ can be a polynomial as suggested in \cite{schobeiri2014applied}:

\begin{equation}\Pi=f_1 I+f_2 D+f_3 D\cdot D\end{equation}

Now, neglecting the higher order term we have:

\begin{equation}\Pi=f_1 I+f_2 D\end{equation}

Since $\Pi$ and $D$ are frame indifferent, $f_1$ and $f_2$ must be invariant. Finding eigenvalues of 2nd order tensors and finding invariants we get:

\begin{equation}I_{1D} = f_1 = Tr(D) = {\bf\nabla}\cdot\bv\end{equation}

\begin{equation}I_{2D} = f_2 = -\frac{1}{4}\Bigg[\Bigg(\frac{\partial u}{\partial y}+\frac{\partial v}{\partial x}\Bigg)\Bigg]^2+\frac{\partial u}{\partial x}\frac{\partial v}{\partial y}\end{equation}

\end{omitext}

\begin{figure}[ht!]
 \centering
 
  \includegraphics[scale=0.6]{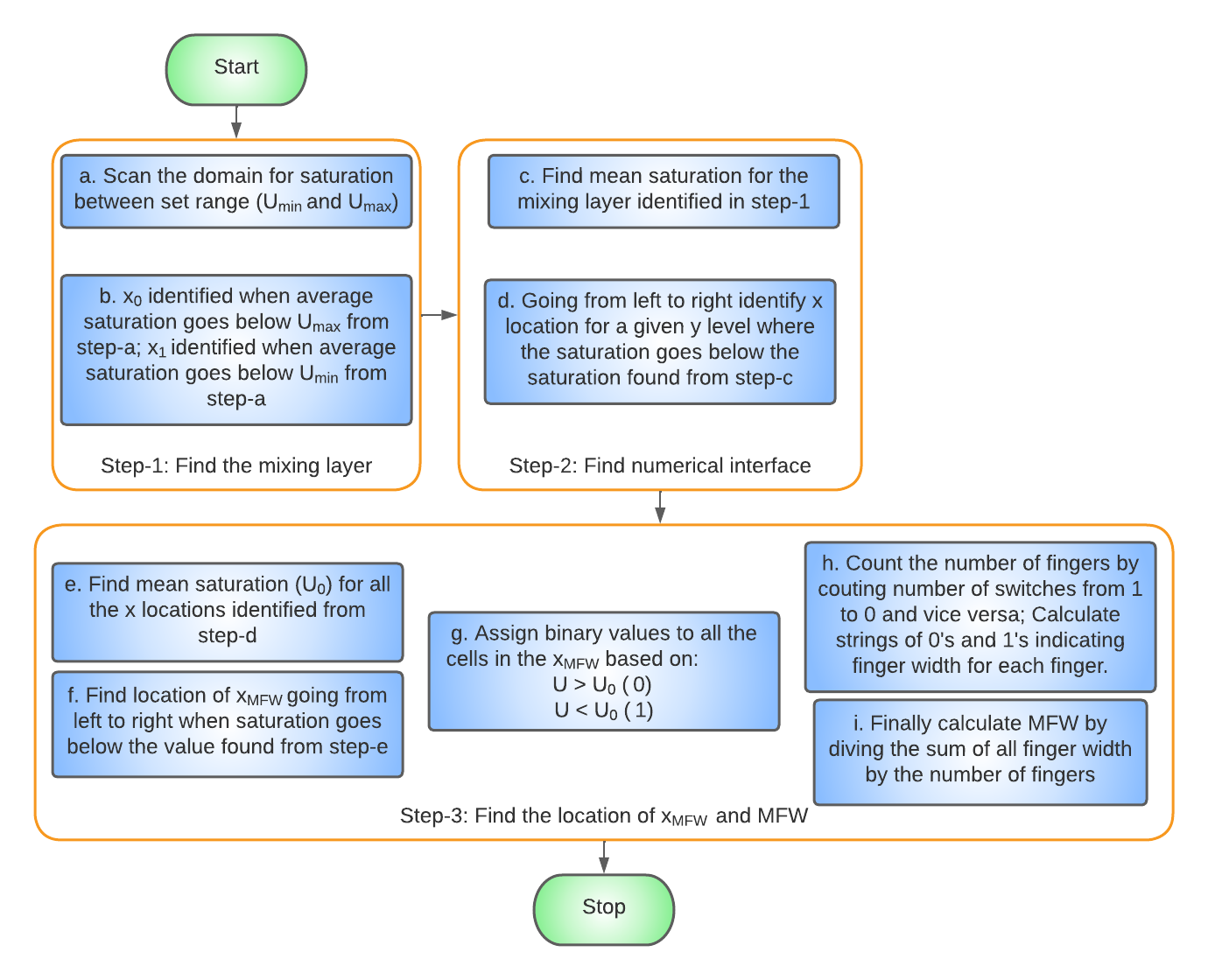}
  \caption{Flowchart for mean finger width calculation}
  
  \label{fig:fig21}
\end{figure}

\begin{figure}[ht!]
 \centering
 
  \includegraphics[scale=0.6]{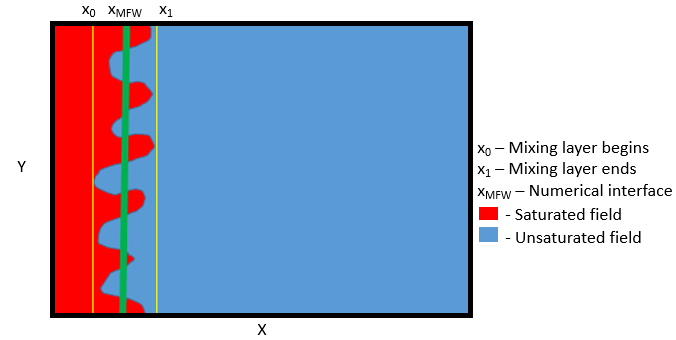}
  \caption{Mixing layer, fingers and numerical interface in rectilinear flooding}
  
  \label{fig:fig22}
\end{figure}

\subsection{Post processing for mean finger width calculation}\label{appendix:fingerwidth-calculation}

This appendix has bearing for section~\ref{section:simulation-in-homo}.
Following procedure is used to calculate the mean finger width (flowchart shown in Fig.~\ref{fig:fig21}) which is based on identifying the mixing layer, fingers and numerical interface shown in Fig.~\ref{fig:fig22}.
\begin{enumerate}
\item Identify the mixing layer: The 2D domain is scanned at each time step to locate a reduction in saturation. For each $y$ level, let $x_0$ denote the start of mixing layer and $x_1$ denote the end of mixing layer. 
\item Find the mean saturation for the mixing layer: The mean saturation is found from the mixing region by taking a mean of saturation at each $y$ level. All the means are then averaged to find a level set based on that cutoff concentration. As the mean concentration goes below this set value, the $x$ value corresponding to it is defined as $x_{\rm MFW}$ (shown in green in Fig.~\ref{fig:fig22}).
\item Find the number of fingers: After establishing the location of the interface, number of fingers across that interface is counted. This is done by assigning binary coefficients to each cell based on whether or not they are higher than the cutoff concentration. A counter keeps track of the number of cells that are present before the cutoff concentration limit is not met at which point the counter is re-initialized to zero to count the next finger. This way we can find the count of fingers in the interface from the number of time this flip happens.
\item Find mean finger width: The mean finger width is found by adding the counter as mentioned in step-3 and dividing that value by the number of flips (number of fingers).
\end{enumerate}
\vspace{0.2truein}

\n {\bf Author Contributions}\
PD lead in the overall planning and formulation of the study outline. PD also developed the theory, algorithms and implemented the algorithms in an in-house code. RM helped in performing the simulations using advanced computing resources provided by Texas A\&M High Performance Research Computing during the semester he took a course on Hydrodynamic Stability with the first author PD.
\vspace{0.2truein}

\n{\bf Funding}\ 
Financial support from the U.S. National Science Foundation through grant DMS-1522782 and from TAMU internal grant T3 from the office of Vice President of Research to the first author PD is gratefully acknowledged.
Portions of this research were conducted with the advanced computing resources provided by Texas A\&M High Performance Research Computing.
\vspace{0.2truein}

\n{\bf Data Availability Statement}\ 
The datasets generated during the current study are available from the corresponding author on reasonable request.
\vspace{0.2truein}

\n{\bf Software Availability}\ The software generated during the current study are available from the corresponding author on reasonable request. The software will be available on GitHub upon publication of this paper.
\vspace{0.2truein}

\n {\bf Conficts of interest}\ The authors declare no confict of interest.

\bibliographystyle{unsrt}
\bibliography{aipsamp}

\end{document}